\documentclass[sigconf]{acmart}
\AtBeginDocument{%
  }

\setcopyright{acmlicensed}
\copyrightyear{2025}
\acmYear{2025}
\acmDOI{XXXXXXX.XXXXXXX}

\acmConference[Conference acronym 'XX]{Make sure to enter the correct
  conference title from your rights confirmation email}{September 28th – October 1st 2025}{Busan, Korea}
\acmISBN{978-1-4503-XXXX-X/18/06}

\usepackage[nolist]{acronym}
\begin{acronym}
    \acro{HCI}{Human-Computer Interaction}
    \acro{CSCW}{Computer-Supported Collaborative Work}
    \acro{CS}{Computer Science}
\end{acronym}
\usepackage[commandnameprefix=always]{changes}
\usepackage{tabularx}
\usepackage{float}

\newcommand{\add}[1]{\begingroup\textcolor{red}{#1}\endgroup}        

\renewcommand{\add}[1]{#1}    

\copyrightyear{2025}
\acmYear{2025}
\setcopyright{acmlicensed}\acmConference[UIST '25]{The 38th Annual ACM Symposium on User Interface Software and Technology}{September 28-October 1, 2025}{Busan, Republic of Korea}
\acmBooktitle{The 38th Annual ACM Symposium on User Interface Software and Technology (UIST '25), September 28-October 1, 2025, Busan, Republic of Korea}
\acmDOI{10.1145/3746059.3747742}
\acmISBN{979-8-4007-2037-6/2025/09}

\begin{document}

\author{Weirui Peng}
\authornote{Both authors contributed equally to this work.}
\authornote{Work done as a visiting researcher at the University of Notre Dame.}
\affiliation{%
  \institution{Columbia University}
  \city{New York}
  \state{NY}
  \country{USA}
}
\email{wp2297@columbia.edu}

\author{Yinuo Yang}
\authornotemark[1]
\affiliation{%
  \institution{University of Notre Dame}
  \city{Notre Dame}
  \state{IN}
  \country{USA}
}
\email{yinooyang@nd.edu}

\author{Zheng Zhang}
\affiliation{%
  \institution{University of Notre Dame}
  \city{Notre Dame}
  \state{IN}
  \country{USA}
}
\email{zzhang37@nd.edu}

\author{Toby Jia-Jun Li}
\affiliation{%
  \institution{University of Notre Dame}
  \city{Notre Dame}
  \state{IN}
  \country{USA}
}
\email{toby.j.li@nd.edu}

\renewcommand{\shortauthors}{Peng et al.}
\newcommand{\ino}[1]{\textcolor{pink}{\textbf{*Yinuo*}: #1}}

\begin{abstract}
Flipped classrooms promote active learning by having students engage with materials independently before class, allowing in-class time for collaborative problem-solving. During this pre-class phase, asynchronous online discussions help students build knowledge and clarify concepts with peers. However, it remains difficult to engage with temporally dispersed peer contributions, connect discussions with static learning materials, and prepare for in-class sessions based on their self-learning outcome. Our formative study identified cognitive challenges students encounter, including navigation barriers, reflection gaps, and contribution difficulty and anxiety. We present \sys{}, an AI-assisted discussion platform for pre-class learning in flipped classrooms. \sys{} helps students identify posts with shared conceptual dimensions, scaffold knowledge integration through conceptual blending, and enhance metacognition via personalized reflection reports. A lab study within subjects (n = 12) demonstrates that \sys{} improves discussion engagement, sparks new ideas, supports reflection, and increases preparedness for in-class activities.

\end{abstract}


\newcommand{\sys}{\textsc{Glitter}}

\title[\sys{}: An AI Platform for Asynchronous Discussion in Flipped Learning]{\sys{}: An AI-assisted Platform for Material-Grounded Asynchronous Discussion in Flipped Learning}





\begin{teaserfigure}
  \centering
  \includegraphics[width=0.80\textwidth]{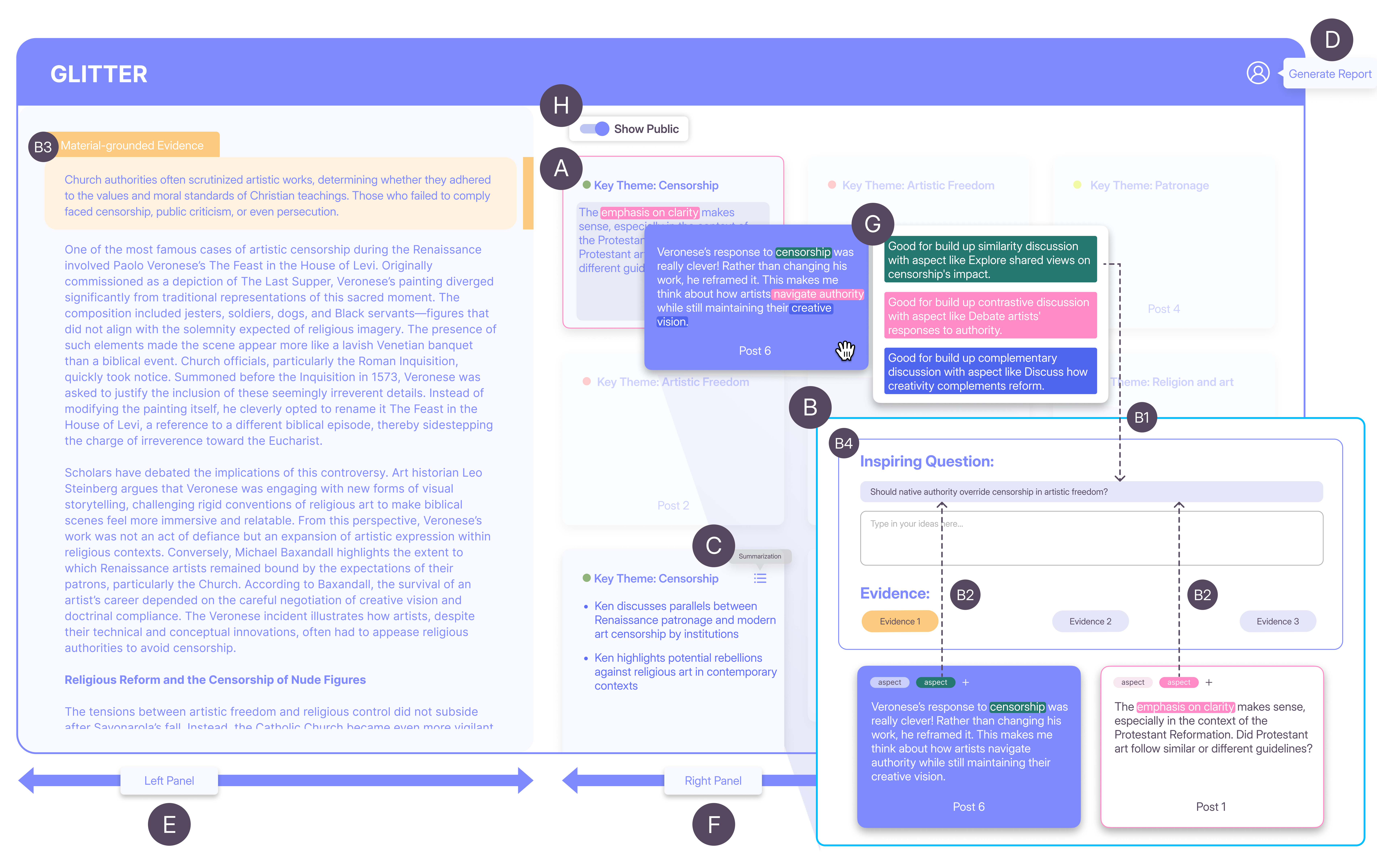}
  \caption{\sys{}'s interface features a left panel for reading (E) and a right panel for discussion (F) that support pre-class learning. After completing their independent reading, students can click the ``Show Public'' button (H) to access and view their peers' discussion contributions. The system enables navigation through posts with shared conceptual affinities (A) and provides content summaries (C) for quick evaluation. When students drag posts toward others' contributions, \sys{} visualizes conceptual connections (G) between ideas. To foster discussions, the blending page (B1) supports the blending selected aspects of the student's posts with their peers' (B2) using the conceptual blending feature (B), which generates inspiring question (B4) and material-grounded evidence (B3). To prepare for in-class discussion, students can access personalized reports (D) showing their learning patterns and participation insights.}
  \label{fig:teaser}
\end{teaserfigure}

\maketitle

\section{Introduction}
Flipped classrooms---where students engage with instructional materials before class and participate in interactive learning activities during class---offer a student-centered alternative to traditional lecture-based instruction~\cite{akccayir2018flipped, betihavas2016evidence,bishop2013flipped}. Unlike traditional lecture-based instruction, this method places emphasis on the \textit{pre-class} phase, which involves deliberate, in-depth engagement with learning materials and meaningful, material-anchored asynchronous discussions~\cite{betihavas2016evidence, lai2016self}. By fostering self-directed learning and peer interaction, flipped classrooms can deepen comprehension, support collaborative sensemaking, and enable iterative refinement of ideas through feedback~\cite{chen2014flip, gilboy2015enhancing, brookfield2012discussion}.

Despite these advantages, students often face significant challenges during self-guided learning and discussions in the pre-class phase. Without sufficient scaffolding, students may struggle to initiate and sustain meaningful asynchronous discussions with peers based on unfamiliar reading materials. This process requires students to sift through peer posts, identify those relevant to their interests, and contribute insightful material-based responses that enhance discussion and promote collaborative learning~\cite{smith2013student, bergmann2012flip}. Moreover, students frequently encounter difficulties in preparing for in-class sessions based on their self-learning outcome and discussion reflection. They tend to forget the material and discussion content they have engaged with, and fail to recognize knowledge gaps from the pre-class phase. These issues undermine the intended synergy between the two phases of the flipped model~\cite{chen2015students, sun2017effect, lai2016self}.


The existing learning platforms used in flipped classrooms, such as Google Classroom~\cite{googleclassroom}, Canvas~\cite{canvas}, and Edmodo~\cite{edmodo}, primarily focus on logistical features like assignment management and offer only basic functions for online collaboration and material learning. For instance, Canvas can integrate with the Perusall~\cite{perusall} reading tool, allowing annotation and commenting on assigned reading materials. While such tools support flipped learning in principle, they do not sufficiently address the cognitive and reflective challenges students face during the pre-class phase.

In a formative study with four students, we identified key challenges in pre-class asynchronous discussions: difficulty formulating meaningful contributions and connecting discussions with learning materials. Students also expressed needs for better navigation to relevant discussion threads and tools to reflect on learning outcomes before class. Based on these findings, we identified four design goals: \textbf{(1) Facilitating efficient exploration and navigation} of discussion space, helping students identify discussions with shared affinity types and comprehend existing discussion; \textbf{(2) Fostering collaborative knowledge synthesis} without overriding human-led knowledge construction, maintaining student-led learning as central; \textbf{(3) Encouraging active and meaningful engagement with broader learning contexts} to promote comprehensive and critical perspectives; \textbf{(4) Fostering metacognitive development} through structured self-reflection on reading comprehension and discussion contributions, helping students monitor their learning and prepare for in-class activities.

To address these needs, we introduce \sys{}\footnote{\textbf{\sys{}} stands for \textbf{G}uided \textbf{L}earning \textbf{I}nterface for \textbf{T}ext-based \textbf{T}hinking, \textbf{E}ngagement and \textbf{R}eflection.}, an AI-assisted platform designed to  enhance asynchronous, material-grounded discussions in the pre-class phase of flipped classrooms.


\sys{} incorporates five intelligent features to support students in meaningfully connecting learning materials with relevant discussion posts based on their individual learning states.
\begin{enumerate}
    \item \textbf{Affinity-based navigation} (Figure \ref{fig:teaser}. A) visually maps conceptual links between posts, enabling students to efficiently identify relevant contributions in complex, content-rich discussion spaces.
    \item \textbf{AI-driven content summarization} (Figure \ref{fig:teaser}. C) reduces cognitive load by distilling lengthy posts into concise summaries, helping students quickly grasp core ideas.
    \add{\item \textbf{Multi-Framework Keyword Highlighting} (Figure~\ref{fig:teaser}. G) reveals discussion pathways between posts through highlighted key terms, helping students identify meaningful connections between their own ideas and peers' perspectives.}
    \item \textbf{Conceptual blending tools} (Figure~\ref{fig:teaser}. B), grounded in Conceptual Blending Theory~\cite{fauconnier1998conceptual}, scaffold collaborative synthesis by helping students merge insights from their own and peers’ posts. The system then generates material-based questions and supporting evidence to guide deeper discussion.
    \item \textbf{Personalized interactive reports} (Figure~\ref{fig:teaser}. D) provide visualizations of students’ reading behavior, discussion contributions, and peer interactions, supporting reflection and identification of knowledge gaps prior to class.

\end{enumerate}


To evaluate \sys{}, we conducted a lab study with 12 participants. The results showed that students effectively used the system's features to improve their pre-class preparation and engagement in discussions. Participants particularly appreciated how the cell-based interface, coupled with AI-powered tools, facilitated active interaction with learning materials, stimulated new ideas during discussions, supported meaningful reflection before class, and enhanced their readiness for in-class activities. We also conducted an exploratory deployment with 21 students in a college classroom, yielding initial insights into how students integrated \sys{} into their regular learning routines.

In summary, this paper presents the following contributions:
\begin{itemize}
\item \sys{}, an AI-assisted discussion platform for pre-class learning in flipped classrooms, enabling semantic post navigation, content summarization, conceptual blending, and metacognitive reflection.
\item A within-subjects lab study with 12 participants validated the usability and effectiveness of \sys{}. The study results suggest \sys{}’s ability to enhance engagement with learning materials, spark new ideas, support pre-class reflection, and improve readiness for in-class activities. The lab study results were also complemented by a lightweight exploratory deployment in a college-level course with 21 engineering students. 
\end{itemize}

\section{Related Work}
\subsection{The Pre-Class Phase in the Flipped Classroom}
\label{sec:flip}
The flipped classroom shifts initial learning to a pre-class phase, where students independently engage with instructional materials (e.g., videos, readings), reserving class time for active problem-solving and collaborative application \cite{bishop2013flipped}. This instructional approach, grounded in active learning principles, has been associated with improved student engagement and enhanced academic performance compared to traditional lecture-based methods \cite{lo2017flipped, akccayir2018flipped}.

Despite its benefits, the pre-class phase poses several cognitive and metacognitive challenges for students. Learners often superficially engage with pre-class materials due to limited motivation or insufficient self-regulation strategies, diminishing the effectiveness of subsequent in-class activities \cite{lo2017flipped, lai2021selfregulated}. Additionally, students frequently experience cognitive overload due to the volume or complexity of assigned materials, which can hinder deep comprehension \cite{abeysekera2015flipped}. The lack of immediate instructor feedback exacerbates these challenges, leaving students uncertain about their level of understanding and unable to effectively address confusions as they arise \cite{zhao2021peer}. Building upon these insights, our approach leverages AI-assisted features to provide material-grounded evidence to facilitate students' comprehension on teaching materials and personalized metacognitive support. Specifically, our system provides learners with personalized reports summarizing their engagement and highlighting areas for deeper reflection, thereby scaffolding their cognitive and metacognitive processes and enhancing overall preparedness for active in-class learning.

\subsection{Material Based Asynchronous Discussions}
Material-based asynchronous discussions anchor interactions directly to learning content, allowing students to collaboratively annotate, comment, and discuss specific segments of texts or videos, thereby enhancing comprehension and community engagement \cite{brush2002webann, cui2024active}. Early systems like WebAnn demonstrated that anchoring discussions to documents can nearly double student participation compared to traditional forums and produce more focused, content-specific dialogue \cite{brush2002webann}. \add{Anchoring discussions in content inherently leverages principles of \textit{collaborative learning}, where students co-construct knowledge through interactive dialogue and shared inquiry. Compared to solitary study, collaborative learning within asynchronous discussions promotes deeper engagement by enabling learners to collectively explore and refine ideas, thus enhancing critical thinking and fostering a robust learning community \cite{stahl2006group, dillenbourg1999collaborative}.} Recently, social annotation platforms (e.g., Perusall) have become increasingly popular in MOOCs and classrooms, allowing students to collectively annotate readings or lecture slides and engage in threaded discussions directly within materials. Empirical studies indicate that such annotation discussions can improve students' preparation and performance; for example, active annotation participation correlates with better post-class outcomes \cite{cui2024active}. Additionally, students report higher motivation and engagement through social annotation compared to solitary activities, fostering a stronger sense of community and shared inquiry. 

However, existing systems remain limited in their ability to scaffold active engagement and support meaningful and in-depth discussions. While some tools offer features such as tagging, summarization, or contextual visualizations to reduce the cognitive load of reading and participation~\cite{zhang2018making, zhang2017wikum, lee2016spotlights, chang2023citesee}, their effectiveness is often constrained by a lack of alignment with students’ learning goals and cognitive states, limiting their ability to foster self-directed engagement. Other platforms, such as NB~\cite{zyto2012successful} and Perusall, enable localized interactions through sidebar annotations but provide limited support for conceptual integration across segments of the material and frequently lead to information overload when texts are densely annotated~\cite{johnson2017social}. As a result, students often struggle to pose open-ended, cross-segment questions or engage in sustained, coherent dialogue, and discussions often remain superficial, centered on immediate clarifications rather than comprehensive synthesis. \add{Comprehensive synthesis refers to students' ability to integrate multiple perspectives, identify overarching themes, and construct coherent mental models from fragmented discussions. Pedagogically, achieving synthesis is crucial because it moves learning beyond simple recall or isolated fact recognition toward deeper cognitive processes such as analysis, evaluation, and creation—aligned with higher-order thinking skills outlined in Bloom's taxonomy \cite{anderson2001taxonomy}. Empirical evidence suggests that learners who engage in synthesizing information from multiple sources demonstrate better conceptual understanding, transferability of knowledge, and improved critical thinking skills, all of which are essential outcomes in active learning environments \cite{suthers2008empirical, chi2014icap}.} Furthermore, although systems like Perusall and Canvas support pre-class preparation, they lack mechanisms for metacognitive monitoring and learning reflection, making it difficult for students to track progress or identify knowledge gaps—ultimately hindering deeper classroom engagement and effective peer collaboration.

\sys{} addresses these limitations by integrating contextual, material-based interactions with enhanced discussion features. \add{The design of \sys{} draws explicitly from collaborative learning theories, particularly knowledge building theory \cite{scardamalia2006knowledge}, which emphasizes community-driven idea improvement and shared knowledge creation. Unlike existing tools such as Perusall and NB, which primarily support localized annotations and commenting within specific text segments, \sys{} enables students to discover conceptual connections across a broader range of contributions through affinity-based navigation that visually represents conceptual connections among posts, helping students identify contributions sharing similar conceptual dimensions and extending beyond immediate textual context to identify thematically related discussions throughout the material. Building on these connections, the system's conceptual blending features enable students to actively merge insights from their own and peers' posts, facilitating collaborative knowledge construction through material-anchored synthesis and evidence-based discussion scaffolding.} 
\sys{} also preserves deep reading through distinct Private and Public modes, ensuring students complete independent material engagement before viewing and responding to peers' contributions. \add{This design is grounded in an established collaborative learning practice, Think-Pair-Share \cite{lyman1987think}, by promoting individual reflection before public discourse, thereby enhancing the depth and quality of collaborative exchanges. Additionally, while systems like Perusall and Canvas support pre-class preparation, they lack mechanisms for metacognitive monitoring and learning reflection. \sys{} addresses this gap through interactive reports that provide metacognitive scaffolds, summarizing students' engagement patterns and highlighting areas for deeper reflection, thereby supporting progress tracking and knowledge gap identification to enhance classroom engagement and peer collaboration.}\looseness=-1

\subsection{Supporting Learning and Thinking Through Conceptual Blending}

\add{Conceptual blending describes how people naturally combine ideas from different domains to create new understanding. For example, when students learn about atomic structure, they might blend their knowledge of the solar system (planets orbiting the sun) with atomic theory (electrons around a nucleus) to form the ``planetary model'' of the atom. This cognitive process of merging distinct concepts into unified mental representations lies at the heart of creative thinking and deep learning \cite{fauconnier1998conceptual}. In educational contexts, conceptual blending could become particularly valuable for asynchronous discussions because it helps students move beyond simply sharing isolated thoughts to actively synthesizing diverse perspectives into richer understanding. When students encounter multiple viewpoints in discussion forums, blending theory provides a framework for identifying meaningful connections and creating integrated insights that wouldn't emerge from individual posts alone.} Conceptual Blending Theory offers a cognitive framework explaining how people integrate elements from different mental spaces to produce new ideas and understandings. Fauconnier and Turner argue that human learning and thinking fundamentally depend on such blending processes \cite{Fauconnier2002Way}, which operate as routine, often subconscious mechanisms in everyday cognition, seamlessly combining disparate knowledge frames into unified mental representations \cite{fauconnier1998conceptual}. This process represents a form of combinational creativity \cite{Boden1998Creativity} that enables constructing metaphors, innovation through analogical reasoning, and abstract problem solving \cite{Veale2019Conceptual}.

 Prior research has demonstrated the benefits of computational tools that support conceptual blending for creative design \cite{Dow2011Prototyping}. VisiBlends \cite{Chilton2019VisiBlends} and VisiFit \cite{Chilton2020VisiFit} proposed novel pipelines for blending visual objects to convey integrated meaning, while PopBlends automatically suggests conceptual blends of reference images \cite{Wang2023PopBlends}. Domain-specific applications include ICONATE \cite{zhao2020iconate} for icon generation through mixing different visual elements, FashionQ \cite{jeon2021fashionq} for fashion design, and Artinter \cite{chung2023artinter} for recombining style elements to facilitate communication. In a related approach, 3DALL-E \cite{liu2023threedalle} offers an ideation system that identifies granular conceptual elements and integrates them to create comprehensive prompts for text-to-image models. Conceptual blending has also informed design methodologies \cite{Imaz2007Blends}, underscoring its value as a framework for interactive tools that augment human reasoning and creativity. In educational contexts, conceptual blending theory explains how learners generate novel insights by merging multiple ideas or analogies. For example, blending has supported student sensemaking in physics by enabling integrated conceptual structures \cite{Odden2021}, helped students creatively engage with complex scientific phenomena through intuitive metaphors \cite{Gregorcic2021}, and facilitated students' integration of mathematics and physics concepts during problem-solving \cite{VanDenEynde2020, Hu2013}. Additionally, blending has been observed in collaborative small-group learning, where students jointly construct integrated mental models through social interaction, deepening their conceptual understanding of challenging topics such as quantum mechanics \cite{Hoehn2018}. Conceptual blending can enrich material-based discussions by offering a structured framework for integrating diverse perspectives with learning materials, helping learners connect their ideas with peers' contributions to foster innovative thinking. By structuring discussions around specific materials, blending can drive students to creatively synthesize diverse ideas grounded in evidence. However, using conceptual blending in this context also poses challenges. Students often struggle to identify meaningful conceptual intersections between their ideas without guidance. Effective scaffolding is therefore essential to help students discover these conceptual connections, providing enough structure to focus their reasoning while still encouraging creative exploration of potential conceptual blends.
 
To address this, \add{drawing on conceptual blending theory as its foundational framework,} \sys{} implements a Multi-framework Keyword Highlighting feature to visually suggest potential connections between posts and a Conceptual Blending with Anchoring Evidence feature to guide students in merging their ideas with those of peers. The system treats students' posts and peers' content as ``input spaces'', facilitating the discovery of novel relationships between seemingly disparate concepts and transforming them into integrated mental models.

\section{Formative Study}

Designing effective support for flipped classrooms requires a grounded understanding of students’ actual experiences with material-based asynchronous discussions. While prior research has demonstrated the potential benefits of flipped learning and explored technical interventions, less attention has been paid to the specific cognitive and participatory challenges students face during the pre-class discussion phase. 

To address this gap, we conducted a formative study to investigate how students engage in asynchronous discussions, the barriers they encounter, and the kinds of support that could enhance their learning experience.

Our study was guided by the following research questions:
\begin{enumerate}
    \item How do students navigate and engage with material-based asynchronous discussions in flipped classrooms?
    \item What cognitive challenges do students face when attempting to connect their own ideas with peers’ contributions?
    \item What difficulties do students encounter when preparing for in-class sessions based on their pre-class learning and discussions?
\end{enumerate}

\subsection{Process}
We conducted semi-structured interviews to explore students' experiences with asynchronous discussions in flipped classroom contexts, with a focus on identifying both strengths and limitations of current practices and tools.  We recruited four participants \add{(detailed information can be found in Table \ref{tableformative} in the Appendix)} through local mailing lists and personal networks. All participants had prior experience with online learning platforms that supported asynchronous discussions in flipped learning environments.

During the interviews, we asked participants to describe how they contributed to material-based discussions, the cognitive difficulties they encountered (e.g., making sense of peer contributions, connecting ideas to course materials), and how these experiences shaped their preparation for in-class sessions.

\add{We conducted an inductive thematic analysis for our formative study. Two researchers independently coded the transcripts, then discussed and reconciled differences to collaboratively develop a shared codebook. Building on this codebook, we engaged in iterative team discussions to synthesize broader, high-level themes.} With this approach, we identified four key themes that reflect the challenges and unmet needs students experience during the pre-class phase of flipped learning. These findings informed the design goals and system features of \sys{}, described in the following sections.

\subsection{Findings}
\subsubsection{Finding 1 (F1): \add{Students identified thematic navigation as a valuable support for locating relevant discussions.}}
All four participants (P1–P4) \add{highlighted the potential value of} navigation tools that connect them to relevant discussion threads within material-based asynchronous discussions. \add{While some comments framed this as a desirable enhancement rather than an essential need, participants consistently emphasized how such features could reduce effort in locating related content and promote collaborative engagement.} For example, P1 mentioned that: \textit{``...combining similar annotations to another (is) a nice feature to have...make me feel I am not the only one who feels this way''} (P1). Participants also noted that using highlights to guide navigation through relevant content would be effective. As P3 explained: \textit{``...some text, like topic sentence and what they want to stress about can be highlight(ed) to help us easily navigate through''} (P3). P2 also noted that having summaries or organized visualizations of discussion threads would significantly reduce cognitive load and help them better connect to broader learning contexts. This illustrates the benefits of effectively facilitating students’ navigation to locate and engage with relevant discussions with shared affinity aspects.

\subsubsection{Finding 2 (F2): Contribution anxiety limits students’ ability to engage in meaningful discussions}
Three of the participants (P2-P4) reported difficulty forming constructive perspectives that contribute meaningfully to material-based asynchronous discussions. P3 expressed frustration with the challenge of proposing new perspectives, particularly when discussions already have numerous replies: \textit{``a lot of people discussing about a typical question actually made me a little bit frustrated...I worry I can't add something new''} (P3). This \add{hesitation} (P4). 

Such anxiety was compounded by the perceived complexity of the material and students’ limited prior knowledge. As P3 pointed out, \textit{``since we didn’t learn the knowledge beforehand, many discussions are vague and difficult to understand''} (P3). To address these challenges, P2 suggested \textit{``tools that connect responses to specific parts of the material could help us better understand discussions and generate meaningful replies''} (P2). To overcome these barriers, participants recommended highlighting related material sections and summarizing ongoing discussions to provide a foundation for meaningful contributions. \looseness=-1

\subsubsection{Finding 3 (F3): Students struggle to connect discussions with broader learning materials}
Participants (P1, P3, P4) reported challenges in connecting discussion threads to broader learning materials, limiting their ability to fully engage with and benefit from asynchronous discussions. For example, as P4 mentioned, \textit{``the material that we're learning in class is related...I want to know broader knowledge connections in the material to build more understanding''} (P4). Participants who had prior experience with platforms that link discussions to specific parts of the material noted that navigating between content sections and discussion threads imposed a significant cognitive load. P1 highlighted the difficulty of identifying related content for deeper learning, while P3 described the process as inefficient and time-consuming: \textit{``it is really time consuming to scroll down the book to find concepts from the material while participating in discussions''} (P3). 

These findings point to the need for tools that more seamlessly link discussion threads with related sections of the learning material, supporting more efficient navigation and integration of knowledge.

\subsubsection{Finding 4 (F4): Students \add{could} benefit from reflective tools that foster metacognitive awareness and support knowledge transfer}
Participants highlighted the value of reflecting on discussion outcomes to reinforce understanding and prepare for in-class sessions. Four participants (P1-P4) emphasized that reflecting on discussion outcomes helps consolidate their understanding of the material and identify key takeaways before engaging in class discussions. For instance, P1 mentioned revisiting past discussions to reinforce their grasp of key concepts, stating, \textit{``Discussing it with others makes it easier to remember key points and recall specific parts of the reading''} (P1). Similarly, P4 expressed a desire to review their own questions and comments, as well as others' replies, to prepare for class—suggesting that such reflections promote confidence and readiness. 

These reflections underscore the value of tools that summarize students’ discussion history, highlight key learning moments, and support metacognitive engagement. By enabling students to monitor their evolving understanding, such tools can foster deeper learning and improve the transition from asynchronous preparation to in-class collaboration.

\subsection{Design Goals}
Drawing on our formative study findings and cognitive theories of learning in flipped classrooms (Section \ref{sec:flip}), we identify four key design goals for an AI-enhanced discussion platform to support pre-class asynchronous engagement:

\begin{itemize}
    \item \textbf{DG1: Facilitate thematic exploration of discussion spaces.} 
Reduce navigation barriers by highlighting conceptual affinities and organizing related content, enabling students to efficiently discover and engage with relevant discussions.

    \item \textbf{DG2: Support collaborative knowledge synthesis while preserving student agency.} 
Provide scaffolding for combining ideas from multiple contributors without overriding students’ interpretive autonomy, ensuring learners remain central to meaning-making processes.

    \item \textbf{DG3: Foster connections between discussions and broa\-der learning materials.} 
Encourage integrative thinking by linking peer discussions to related concepts across the curriculum, deepening understanding through cross-contextual connections.

    \item \textbf{DG4: Scaffold metacognitive reflection and knowledge monitoring.} 
Promote self-awareness and preparation through structured reflection on reading comprehension and discussion contributions, helping students track learning progress and identify areas for growth ahead of in-class activities.

\end{itemize}

\section{The \sys{} System}
\sys{} is a cell-based discussion platform designed to enhance pre-class material-centered  asynchronous discussions in flipped classroom settings. \sys{} provides key functionalities including conceptual affinity navigation, conceptual blending, content summarization, and personalized report. These features aim to reduce cognitive barriers, foster collaborative knowledge construction, and promote metacognitive awareness. 

Below, we detail the core features and their implementation in \sys{}, followed by an illustrative scenario demonstrating how users use them.\looseness=-1


\subsection{Example Scenario}

Alex is a college student enrolled in a flipped course on climate policy. Each week, he reviews assigned readings and participates in asynchronous discussions prior to class. While he appreciates the flexibility of self-paced learning, he often struggles to navigate fragmented discussions, synthesize diverse perspectives, and connect conversations with core materials.

To address these challenges, Alex uses \sys{}. He begins in Private Mode, allowing him to read and annotate the material independently before viewing others' contributions. While reading, he adds annotations that appear as discussion blocks in the right panel (Figure \ref{fig:teaser}. F). After forming his initial ideas, he clicks `Show Public' button (Figure \ref{fig:teaser}. H) to transition to Public Mode and explore his peers' perspectives.

\begin{figure}[t!]
  \centering
  \includegraphics[width=\columnwidth]{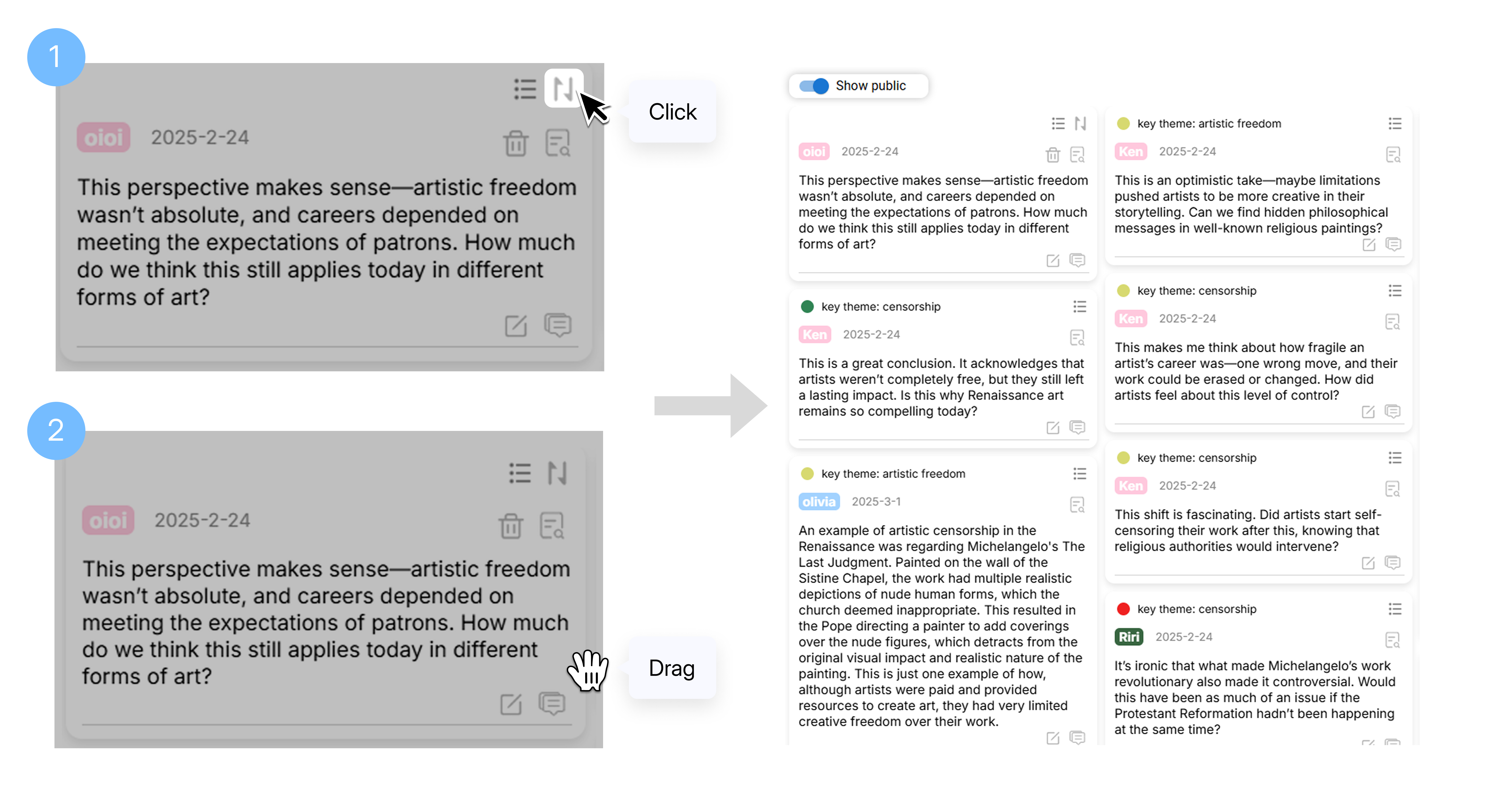}
  \caption{Illustration of Conceptual Affinity Navigation. Students can activate this feature by clicking the ``Order'' button (1) in the top right corner of the post or dragging the post (2). Once triggered, \sys{} identifies shared affinity dimensions between the selected post and all other contributions, assigns each post a concise affinity type, and displays color-coded visual indicators in the top left corner of each post to indicate content relevance.}
 
  \label{fig:affinity_ordering_1}
\end{figure}

When exploring others' contributions, Alex is particularly interested in posts related to his annotation on economic nationalism. He clicks the `Order' button (Figure \ref{fig:affinity_ordering_1}. 1), prompting \sys{} to identify and display affinity types between posts. For Amy's post about the prisoner's dilemma, the system suggests ``Economic Theory Application'' as the shared affinity type, with a green circle indicating high semantic relevance. Intrigued, Alex clicks `Generate Summary' button (Figure \ref{fig:Summary}) for an overview of Amy's post. When hovering over her post, \sys{} highlights key terms in both posts and suggests potential discussion points (Figure \ref{fig:Blending}. a). Intrigued by the connection, Alex navigates to the discussion overview page, which offers targeted suggestions for building on Amy’s ideas.

To deepen the conversation, Alex enters the Blending Page (Figure~\ref{fig:Blending}. b), where the system prompts him with key concepts that potentially connect both his and Amy’s posts. Interested in the theoretical connections, Alex drags Amy's ``Game Theory Dynamics'' and his own ``Economic Nationalism'' aspects into the blending area. This combination prompts \sys{} to generate an Inspiring Question (Figure \ref{fig:Blending}. c1): ``How can international frameworks address both the prisoner's dilemma and economic nationalism to foster climate cooperation?''

To further support his thinking, the system retrieves three relevant, material-grounded evidence blocks aligned with these concepts (Figure \ref{fig:Blending}. c2). Drawing on the suggested question and supporting content, Alex composes a thoughtful response that integrates economic theory and policy, thereby initiating a deeper, interdisciplinary conversation with Amy.

After engaging with materials and discussions, Alex clicks 'Generate Report' button (Figure \ref{fig:teaser}. D) to view an analysis of his learning activities (Figure \ref{fig:Report}). The report visualizes his reading behavior, discussion contributions, and peer interactions, helping him reflect on his engagement and identify knowledge gaps. This process strengthens his preparation for in-class collaboration.

\subsection{Key Features}
\sys{} integrates five key features that support students in pre-class, material-based asynchronous discussions: conceptual affinity navigation, content summarization, multi-framework keyword highlighting, conceptual blending with evidence anchoring, and interactive discussion reports. These features collectively address the cognitive and metacognitive challenges identified in our formative study and align with the design goals outlined earlier.

\subsubsection{Conceptual Affinity Navigation for Discovering Connected Ideas}
Navigating fragmented or disconnected discussions poses a significant challenge for students seeking to build coherent understanding (DG1). To address this challenge, \sys{} implements a conceptual affinity navigation system, which surfaces thematically related posts to guide students toward relevant and meaningful discussions. \add{Anchored in collaborative-learning frameworks, this feature reduces navigation overhead and supports Think-Pair-Share, enabling students to move from Active selection toward Constructive integration of ideas \cite{lyman1987think, chi2014icap}.}

When a student activates the ordering feature---either by clicking the 'Order' button on a post or by dragging a post (Figure \ref{fig:affinity_ordering_1})---\sys{} uses an LLM to perform multi-dimensional analysis evaluating conceptual alignment between the selected post and all other contributions in the workspace. The system identifies shared affinity dimensions and assigns each post an affinity type that articulates the specific conceptual bridge between the student's post and other contributions, establishing common ground across different discussion threads. 

To help students quickly gauge content relevance at a glance, the system implements color-coded visual indicators: green highlights for highly relevant content, yellow for moderately related material, and red for minimally connected contributions. This intuitive visual encoding reduces cognitive burden by allowing students to effectively identify which posts contain substantively related content without requiring detailed reading of each contribution.

This approach prioritizes meaningful conceptual connections while reducing cognitive load by creating intuitive pathways through the discussion space. By exposing the conceptual dimensions that connect seemingly disparate posts and providing supplementary relevance cues, \sys{} enables students to engage more purposefully with discussion content. This facilitates deeper exploration of interconnected ideas across different threads while helping students navigate beyond surface-level similarities to discover richer conceptual relationships.

\begin{figure}[t!]
  \centering
  \includegraphics[width=\columnwidth]{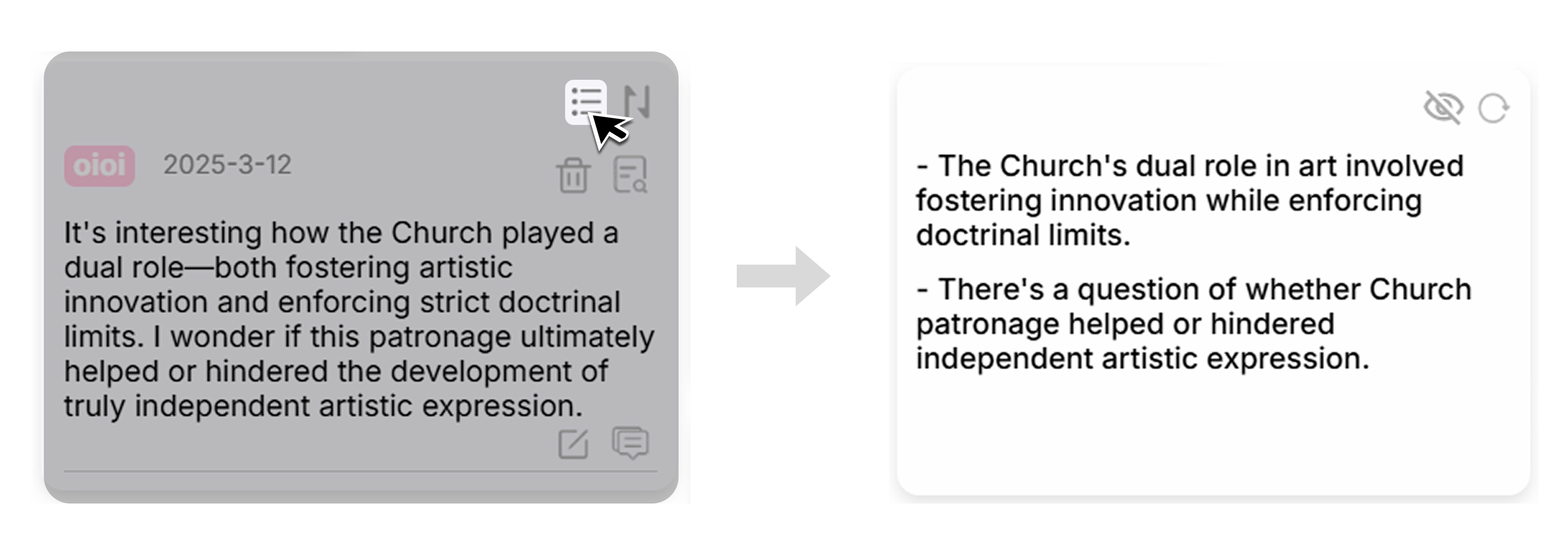}
  \caption{Illustration of Content Summarization. Students can activate this feature by clicking the 'Generate Summary' button. Once activated, \sys{} creates a concise bullet-point summary of the selected post's content.}
 
  \label{fig:Summary}
\end{figure}

\subsubsection{Content Summarization for Efficient Comprehension}
Once students identify relevant posts, understanding and synthesizing multiple contributions can be cognitively overwhelming---especially when posts are lengthy or complex (DG1). After locating posts with shared affinity types, students often face information overload when attempting to process and synthesize diverse perspectives. To address this cognitive burden, \sys{} incorporates a summarization feature designed to condense essential information from the discussion contributions.

When a student encounters a relevant post through the affinity visualization system, they can activate the summarization function with a single click on 'Generate Summary' button (Figure \ref{fig:Summary}). \sys{} then generates a concise yet comprehensive summary that captures the core arguments, evidence, and conceptual connections present in the original contribution. \add{Beyond individual posts, \sys{} also enables students to summarize multiple posts within a thread. By aggregating related contributions in threaded discussions, students can obtain more comprehensive summaries that capture the evolution of ideas and multi-faceted perspectives on complex topics.} This functionality reduces cognitive load by distilling lengthy or complex posts into their essential components, allowing students to quickly grasp key ideas without sacrificing critical nuance. By streamlining information processing, \sys{} enables students to engage more deeply with the content itself rather than expending cognitive resources on simply deciphering and organizing peer contributions.

 

\subsubsection{Multi-Framework Keyword Highlighting for Facilitating Conceptual Connections} Effective collaborative knowledge construction requires students to identify meaningful conceptual relationships between their own ideas and those of their peers (DG2). However, in asynchronous environments, students struggle to identify meaningful connections between their own ideas and their peers' diverse perspectives. To address this challenge, \sys{} implements an interactive keyword visualization system that dynamically reveals potential discussion pathways between contributions.

\begin{figure*}[t!]
  \centering
  \includegraphics[width=\linewidth]{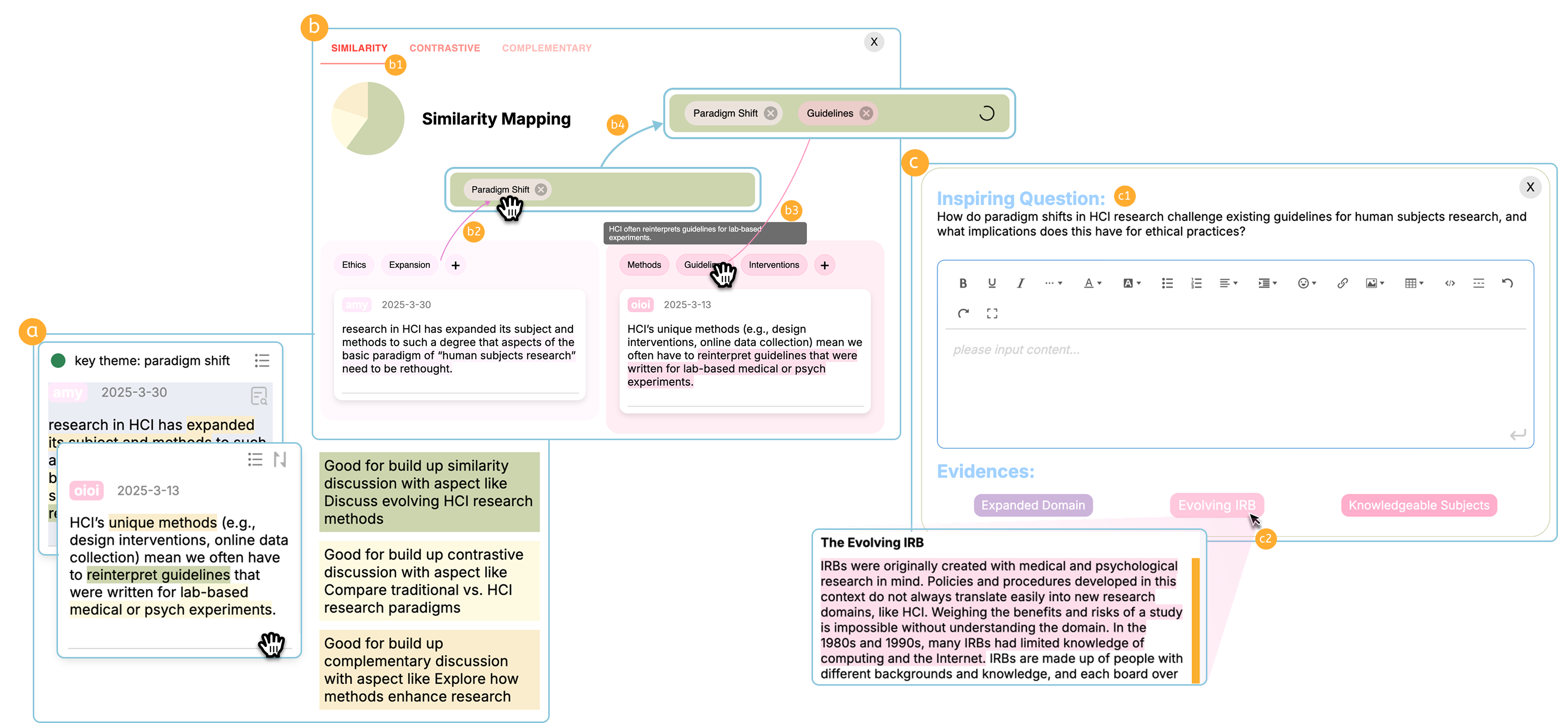}
  \caption{Illustration of the processes of Multi-Framework Keyword Highlighting and Conceptual Blending with Evidence Support. Student drags their post and hovers over another post
   that interests them, \sys{} highlights key terms in both posts that serve as foundations for meaningful discussion development (a). On the blending page, students can first choose a discussion framework (b1) and select one key aspect from each post for knowledge synthesis (b2, b3, b4). Upon selection, \sys{} generates an ``Inspiring Question'' (c1) that bridges the chosen perspectives while simultaneously retrieving three relevant supporting evidence excerpts directly from course materials (c2). These evidence blocks appear color-coded below the text editing box, with each excerpt visually linked to its original context in the learning materials.}
 
  \label{fig:Blending}
\end{figure*}

When a student drags their post and hovers over another post that interests them, \sys{} activates its conceptual connection analysis (Figure \ref{fig:Blending}. a). This direct manipulation interaction follows the HCI principles of explicit user intent \cite{shneiderman1997direct}---dragging mirrors the cognitive act of bringing ideas together while hover feedback enables low-commitment exploration, reducing the cognitive load of comparing multiple posts. The system identifies and highlights key terms in both posts that serve as foundations for meaningful discussion development. \sys{} organizes these connections according to three distinct conceptual frameworks \cite{fauconnier1998blending}—similarity, contrastive, and complementary relationships—each represented through different visual indicators.


By providing these visual cues and targeted suggestions, \sys{} establishes a foundational understanding of how specific aspects of the student's thinking relate to their peer's perspective. This preliminary mapping enables students to recognize which elements of their post would most productively engage with the other contribution, creating a cognitive scaffold for more sophisticated discussion development.

\subsubsection{Conceptual Blending with Evidence Anchoring for Integrated Knowledge Construction}
Meaningful knowledge construction in collaborative learning environments extends beyond simple information exchange, involving a balance between two aspects: the purposeful integration of diverse viewpoints into coherent conceptual frameworks (DG2), and the need for effective discussion contributions to remain \textit{grounded} in course materials to ensure relevance and accuracy (DG3).

Achieving this balance is often challenging for students, particularly when locating supporting evidence across lengthy and complex readings. Manually scanning documents for relevant content imposes a high cognitive load, diverting attention from higher-order reasoning and synthesis.

To address these dual challenges, \sys{} incorporates a conceptual blending mechanism that facilitates the synthesis of ideas across contributions while automating the retrieval of material-based evidence to support discussion development. \add{Following the knowledge-building “rise-above” principle, this mechanism steers learners beyond isolated posts toward collaboratively constructing an evidence-anchored synthesis \cite{scardamalia2006knowledge}.}

When students move from initial keyword exploration (discussion overview phase) to the conceptual blending page, \sys{} provides a structured blending interface (Figure \ref{fig:Blending}. b), where they can select a key aspect from each post for knowledge synthesis. These key aspects---each representing a core thematic element automatically extracted from the post's content---serve as conceptual anchors for the blending process. Each post displays three distinct aspects that encapsulate different dimensions of the contributor's thinking, allowing for targeted integration of specific ideas rather than overly general connections.

Once aspects are selected, the system activates the conceptual blending feature, simultaneously processing the selected aspects, complete post contents, and course materials. \sys{} generates an ``Inspiring Question'' that bridges the selected perspectives and synthesizes these inputs to craft a discussion prompt specifically designed to illuminate connections between the chosen aspects. Concurrently, the system employs Retrieval-Augmented Generation (RAG) techniques to identify three supporting evidence directly extracted from the learning materials. The RAG implementation creates a comprehensive knowledge base exclusively from the provided course materials, ensuring that all retrieved content originates solely from the learning corpus. By constraining the retrieval process to this closed set of materials, the system fundamentally prevents hallucination since the model can only access and reference content that exists within the course materials. These support evidences---often from different sections of the materials that students might not independently connect---appear as color-coded blocks beneath the text editing box, each visually linked to its original context through matching highlights in the source material. This unified approach ensures the Inspiring Question and supporting evidence share the same contextual foundation, creating a coherent discussion framework. The RAG mechanism extracts evidence directly from learning materials, preventing hallucination and maintaining strict grounding in the course materials \add{through this closed-corpus approach}.

By augmenting the evidence retrieval process across extensive materials, \sys{} expands the contextual foundation available for discussion, enabling students to draw connections across broader content domains without the cognitive burden of exhaustive manual searching. This feature reduces the barriers to knowledge integration while simultaneously enriching connections to the broader learning context.

\begin{figure*}[t!]
  \centering
  \includegraphics[width=\linewidth]{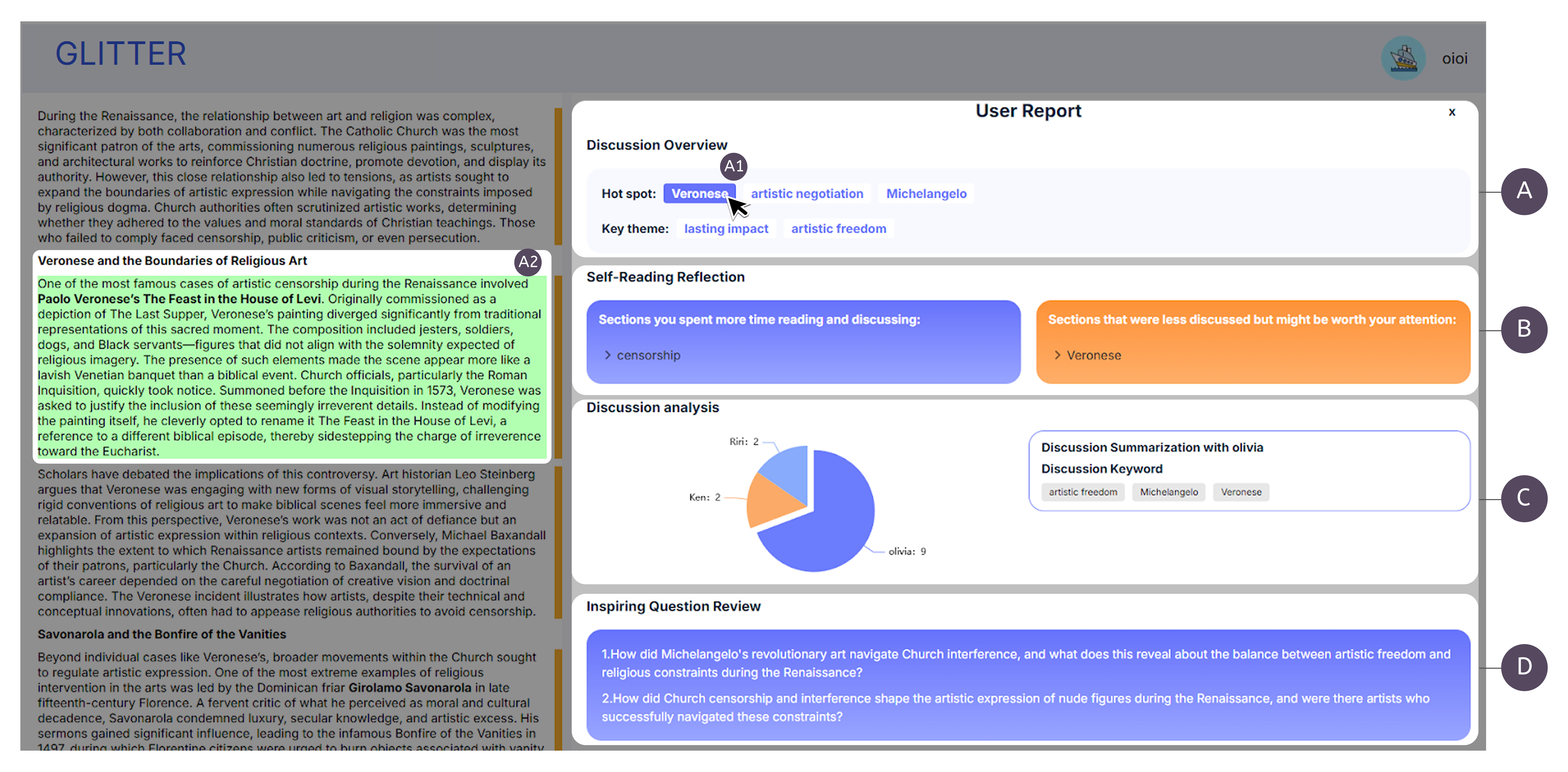}
  \caption{\sys{}'s Personalized Learning Report visualizes students' learning activities through multiple interactive components. The Discussion Overview (A) displays content ``Hot spot'' (A1) that, when hovered over, highlight corresponding material in the reading panel (A2) while identifying recurring discussion key themes. The Self-Reading Reflection section (B) presents frequently engaged content and underexplored content, helping students identify knowledge gaps. The Discussion analysis section contains an interactive pie chart (C) that categorizes interactions with different peers, promoting awareness of collaboration patterns. Students can also review previously generated Inspiring Questions (D) to reinforce key concepts and discussion prompts.}
 
  \label{fig:Report}
\end{figure*}

\subsubsection{Interactive discussion reports for enhancing metacognitive awareness}
To facilitate students' reflection on their learning engagement during the pre-class phase and prepare them for in-class participation (DG4), \sys{} generates personalized interaction reports that visualize their reading and discussion activities (Figure \ref{fig:Report}). 

These reports are designed to promote metacognitive awareness by helping students monitor their activity, identify knowledge gaps, and connect their engagement with specific learning outcomes.

The report features a Discussion Overview component that reveals ``Hot spot'' (frequently discussed content areas) and recurring discussion key themes. When students explore these elements, the system highlights corresponding sections in the original materials, enabling efficient tracing of discussion contexts to source content. 

The report provides a Self-reading Reflection component that displays reading patterns, highlighting both frequently engaged content areas and underexplored sections. For frequently engaged content areas, students can click the summarized theme to access their previous comments and click the light bulb button to receive tailored suggestions for in-class discussion participation. They can also click on the underexplored section to explore materials they are not familiar with, encouraging more balanced engagement across all learning resources.

\sys{} also synthesizes students' discussion activities into an interactive pie chart that categorizes their interactions with different peers. Each segment corresponds to interactions with specific classmates, and clicking any segment reveals detailed discussion summaries along with suggestions for incorporating these insights into future in-class discussions. Students can further filter these exchanges using keyword buttons to quickly locate specific conversation threads.

Together, these components support a range of in-class learning goals when students transition from the pre-class phase to in-class learning:

\begin{itemize}
    \item The \textbf{Discussion Overview} prepares students for group dialogue by surfacing shared topics and divergent perspectives.
    \item The \textbf{Reading Reflection} equips them for analytical activities by revealing both strengths and gaps in content engagement.
    \item The \textbf{Peer Interaction Chart} enables more intentional collaboration by identifying discussion partners with shared or complementary interests.
\end{itemize}

By making learning behaviors visible and actionable, the \sys{} report empowers students to take ownership of their learning process. It serves as both a diagnostic tool and a strategic guide, promoting deeper reflection and more effective preparation for synchronous in-class activities.




\subsection{Implementation}
\label{sec:implementation}

\sys{} is implemented as a web-based application designed for multi-user interaction. It consists of a React-based\footnote{\href{https://react.dev/}{https://react.dev/}} frontend client, a Node.js-based\footnote{\href{https://nodejs.org/en}{https://nodejs.org/en}} backend server, and a MongoDB\footnote{\href{https://www.mongodb.com/}{https://www.mongodb.com/}} database. The AI-powered features are implemented using OpenAI's GPT-4o API\footnote{\href{https://platform.openai.com/docs/models/gpt-4o}{https://platform.openai.com/docs/models/gpt-4o}}, with LangChain\footnote{\href{https://www.langchain.com/}{https://www.langchain.com/}} providing the vector storage capabilities for the retrieval-augmented generation features. The technical implementation details for each key feature are discussed in the following sections. Detailed prompts for each feature are included in the Appendix.

\subsubsection{Conceptual Affinity Navigation}

\sys{} uses the GPT-4o model to analyze the conceptual relationships between the discussion posts. When a student activates the ``Order'' feature, the model evaluates conceptual alignment between their selected post and all other contributions, producing a multi-dimensional analysis that includes relevance scores (0-1 scale) and shared affinity dimension bridges. The system organizes posts through their shared affinity dimensions while providing a visual indication of the strength of the semantic relationship. Each post pair is assigned a concise affinity type (1-2 words, e.g., ``theoretical applications'' or ``practical examples'') that articulates the specific conceptual bridge between them. The system translates semantic proximity into a color-coding scheme—green for high relevance (relevance score over 0.7), yellow for moderate (0.4-0.7), and red for minimal (below 0.4).

\subsubsection{Content Summarization for Efficient Comprehension}

For the discussion summarization feature, \sys{} uses the GPT-4o model to distill complex posts into concise but comprehensive bullet points. The model analyzes both the main content and any nested replies, extracting core arguments and conceptual insights. Our implementation constrains each summary to 1--3 bullet points of no more than 30 words each, with the exact number adapting to content length and complexity. 

\subsubsection{Multi-Framework Keyword Highlighting}

\sys{} implements its multi-framework keyword highlighting feature that dynamically reveals potential discussion pathways between contributions using GPT-4o to analyze both discussion posts through three distinct conceptual lenses: similarity-based discussion (shared perspectives), contrastive discussion (different viewpoints), and complementary discussion (mutually enhancing ideas). \add{The system prompts the model to analyze two discussion posts and calculate a percentage score for each discussion style, indicating the potential for discussion under that particular framework, with all three scores summing to 100\%. For each relationship type with a non-zero percentage, the model extracts brief quotes (1--3 words) directly from the original post content and provides a concise discussion aspect (1--10 words) to guide potential conversations.} These elements are presented through a color-coded visual system: similarity relationships are colored green, contrastive are colored yellow, and complementary are colored orange.


\subsubsection{Conceptual Blending with RAG Implementation}

The conceptual blending feature combines several AI techniques. First, it utilizes GPT-4o model to extract three key aspects from each post, representing core thematic elements of the content. For the evidence retrieval component, it uses a Retrieval-Augmented Generation (RAG) approach that processes the learning materials into text chunks and creates vector embeddings using OpenAI's text-embedding-ada-002 model. These embeddings are stored in a vector database for efficient similarity search.

When a user selects aspects for blending, the system formulates a query based on the selected aspects and post content, retrieves the most relevant text chunks from the vector database, and passes these to GPT-4o along with instructions to generate an Inspiring Question and identify supporting evidence passages. \add{For Inspiring Question generation, users first select one of three predefined discussion styles: similarity-based discussion (shared perspectives), contrastive discussion (different viewpoints), and complementary discussion (mutually enhancing ideas). The system then instructs the model to generate a thought-provoking question (20--30 words) that aligns with the user-selected discussion style and directly relates to the selected aspects and post contents. For evidence retrieval, we require the model to identify three pieces of evidence from the learning materials, with each evidence consisting of a key concept (1--2 words), exact text extraction with original formatting preserved, and clear connections to the discussion question.} This approach ensures that all evidence is directly extracted from the learning materials rather than fabricated, preventing hallucination while maintaining a strict grounding in the content of course materials.

\subsubsection{Interactive Discussion Reports}

The report generation system implements a comprehensive analysis of user interactions using GPT-4o to transform raw engagement data into actionable insights. The analytical pipeline processes three primary data sources: discussion contributions, peer interactions, and engagement with original learning materials to create personalized learning analytics.

The Discussion Overview component identifies ``Hot spot'' of collective attention across learning materials, mapping discussion density to specific content sections. Each ``hot spot'' is assigned a descriptive keyword and linked to its source, facilitating efficient navigation through the discussion landscape.

The Self-Reading Reflection component analyzes individual engagement patterns, contrasting areas of deep engagement with underexplored content regions. This analysis presents both strengths in a student's current engagement and opportunities for more balanced content exploration, encouraging comprehensive coverage of learning materials.
The Peer Interaction analysis visualizes the distribution of discussions across different classmates, highlighting thematic patterns and generating strategic recommendations for leveraging these insights on in-class learning activities. The system presents these insights through interactive visualizations that respond to user exploration, revealing contextual information and adaptive guidance.


\section{Evaluation}
To evaluate the usability, effectiveness, and usefulness of \sys{}, we conducted a lab user study\footnote{The study protocol was reviewed and approved by the IRB at our institution.} with 12 participants. The study aims to answer the following research questions:

\begin{itemize}
    \item \textbf{RQ1}: How does \sys{} impact students' active engagement with discussions?
    \item \textbf{RQ2}: To what extent does \sys{} support idea generation in asynchronous learning environments?
    \item \textbf{RQ3}: How effectively does \sys{} enhance students' meta-cognitive awareness and preparation for in-class activities?
    \item \textbf{RQ4}: What impact does \sys{} have on students' cognitive load when processing information and participating in asynchronous discussions?
\end{itemize} 
\subsection{Lab User Study}
\subsubsection{Recruitment.}
We recruited 12 university students from diverse academic backgrounds (detailed information can be found in Table \ref{tablelab} in the Appendix) who had prior experience with material-based asynchronous discussions and flipped classrooms. The participants included 5 males and 7 females, representing diverse disciplines: 8 in engineering, 2 in social sciences, and 2 in natural sciences.

\subsubsection{Task.}
We designed a 30-minute simulation task to emulate student engagement in asynchronous components of a flipped classroom. The task was structured to reflect a typical pre-class learning cycle, including individual reading, contribution of original comments, engagement in asynchronous discussion, and pre-class reflection. 

To ensure content diversity and authenticity, we selected three college-level readings (approximately 1,000 words each) on art, history, and climate change, sourced from real college-level classes. The participant encounters one of the three readings in a task. To reflect the intellectual breadth of liberal arts education, these materials were selected for their epistemic complexity, disciplinary range, and inclusion of socially and ethically contested topics, providing opportunities for interpretive reasoning and critical peer discussion in flipped learning. 

To balance ecological validity with experimental control, each reading was accompanied by 25 pre-generated discussion posts created using the GPT-4o model; all participants engaged with the same set of posts. This setup allowed us to systematically examine how system features influenced students' interaction with material and peer discourse across consistent content.

\subsubsection{Conditions.}
The user study used a within-subjects design with two conditions:
\begin{itemize}
    \item{\textbf{Baseline}:}
 A simplified version of \sys{} with the material reading and basic discussion features, providing material highlighting, post merging, post reply features but does not support the post affinity analysis, post blending and inspiring question features or the generation of reflection reports.
 \item{\textbf{\sys{}}:} The full version of the \sys{} system including all features.
\end{itemize}

\subsubsection{Study procedure.}
Each participant took part in two task sessions, one using the baseline system and the other using the \sys{} system, following a within-subjects design. To counterbalance the order of conditions and tasks, participants were randomly assigned to different task-condition pairings. 

At the beginning of each task session, participants watched a tutorial introducing the features of the assigned system (around 10 minutes). To ensure a thorough understanding, participants were given the opportunity to try out the system's features using a sample tutorial material.

Following the tutorial, participants completed a guided simulation task designed to mimic a flipped classroom scenario. They first read the assigned material in private mode and created at least two initial comments (10 minutes), and then participated in an asynchronous discussion with existed posts using the system's features (10 minutes). 

Following this, participants engaged in a simulated pre-class reflection task, during which they read the provided materials and discussion content in preparation for upcoming in-class activities (e.g., group discussions, posing questions to the instructor).

After each task session, participants completed a short questionnaire on their experience with the system.

Upon completing both sessions, participants participated in a 10-minute semi-structured interview. The interview explored their overall impressions and comparison of the two systems, perceived benefits, and feedback on specific features. All study sessions were conducted via Zoom and were recorded with participants’ consent.


\begin{figure*}[t!]
  \centering
  \includegraphics[width=\linewidth]{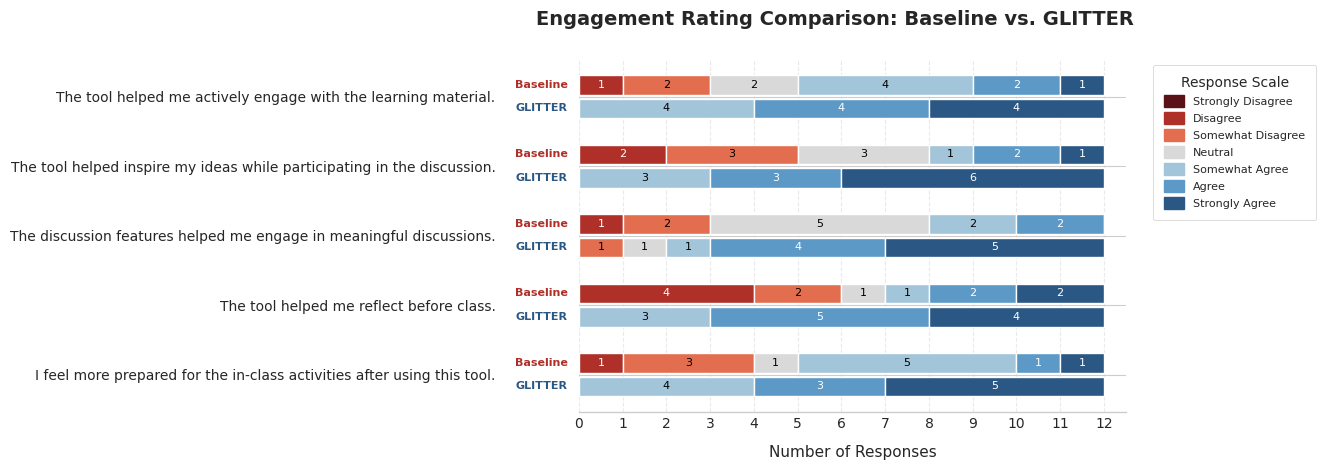}
  \caption{Self-reported questionnaire results about the effectiveness of \sys{} in facilitating student engagement in the lab study}
 
  \label{fig:study_result1}
  \vspace{-3mm}
\end{figure*}

\begin{figure*}[t!]
  \centering
  \includegraphics[width=\linewidth]{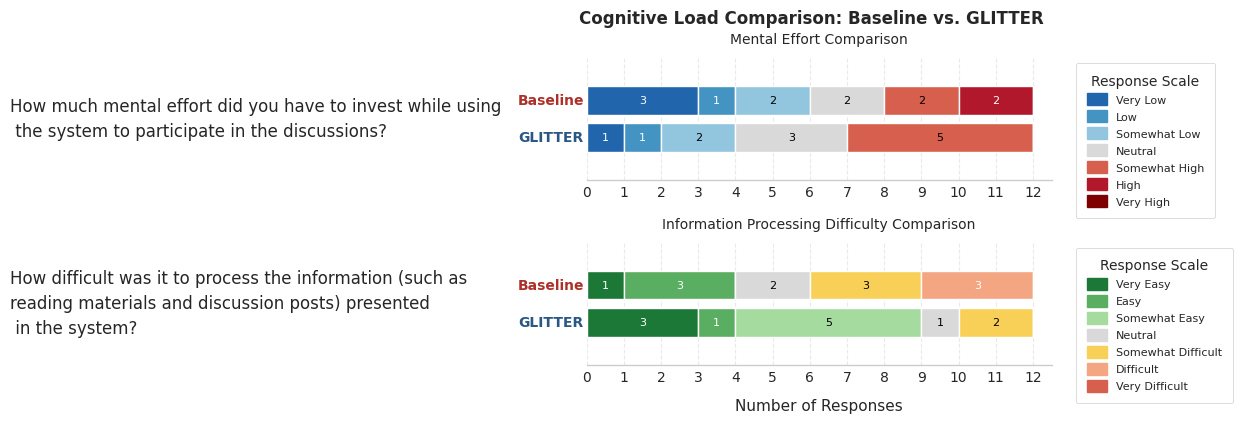}
  \caption{Self-reported results about mental effort and information processing difficulty of using \sys{} in the lab study}
 
  \label{fig:study_result2}
  \vspace{-3mm}
\end{figure*}

\begin{figure*}[t!]
  \centering
  \includegraphics[width=\linewidth]{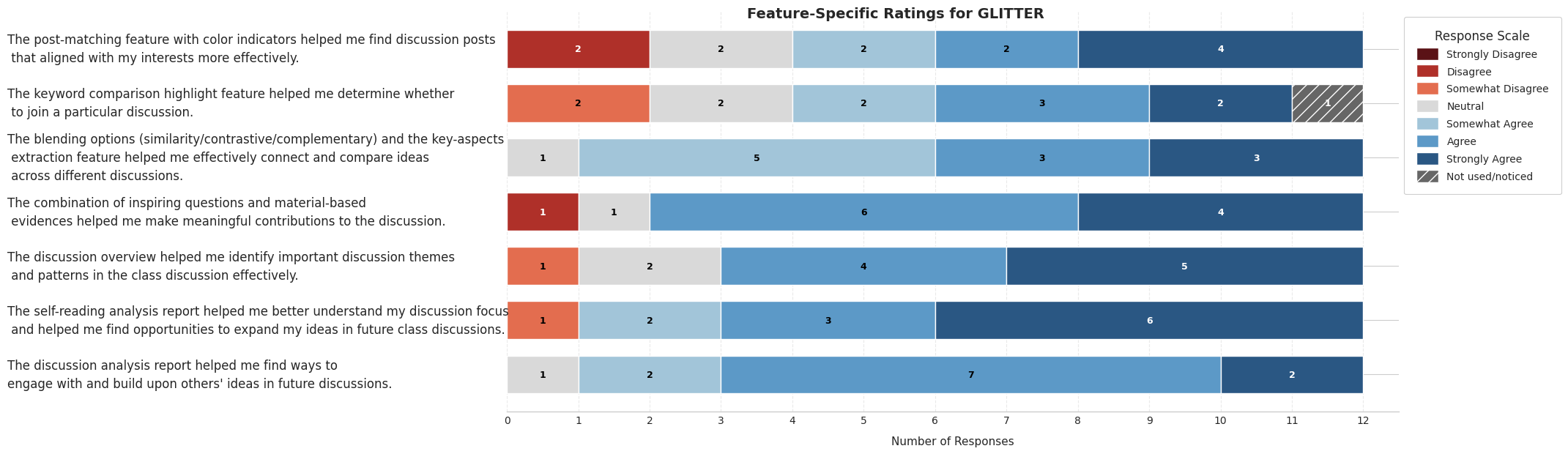}
  \caption{Self-reported results about the usefulness of specific features of \sys{} in the lab study}
 
  \label{fig:study_result4}
  \vspace{-3mm}
\end{figure*}

\begin{figure*}[t!]
  \centering
  \includegraphics[width=\linewidth]{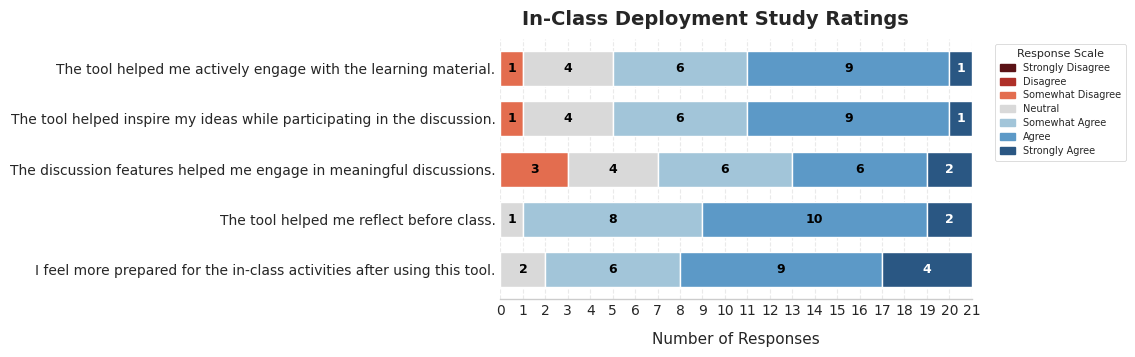}
  \caption{Self-reported results about usability, usefulness and effectiveness of \sys{} in the exploratory deployment}
  \label{fig:study_result5}
  \vspace{-3mm}
\end{figure*}

\begin{figure*}[t!]
  \centering
  \includegraphics[width=\linewidth]{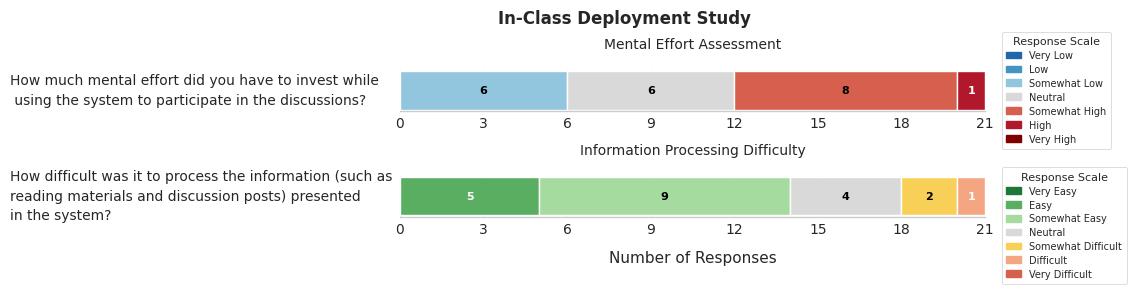}
  \caption{Self-reported results on mental effort and information processing difficulty of using \sys{} in the exploratory deployment}
  \label{fig:study_result6}
  \vspace{-3mm}
\end{figure*}

\subsection{Results}
\subsubsection{Quantitative Results.}
Table~\ref{tab:system_baseline_comparison} presents a comparison between the \sys{} and Baseline methods in terms of the average number of posts made and the average time spent (in minutes). \sys{} shows a significantly higher average number of posts (6) compared to Baseline (4.25) ($p = 0.004 < 0.05$, Wilcoxon signed-rank test). \sys{} also required significantly more time on average (23.17 minutes) compared to Baseline (15.5 minutes; $p = 0.006 < 0.05$, Wilcoxon signed rank test). This suggests that while \sys{} encourages greater user activity, it also result in more time spent on the engagement with materials and interactions.

\begin{table}[h]
\centering
\begin{tabular}{lcc}
\hline
\textbf{Method} & \textbf{Avg Number of Posts} & \textbf{Avg Time (mins)} \\
\hline
\sys{} & 6 & 23.17 \\
Baseline & 4.25 & 15.5 \\
\hline
\end{tabular}
\caption{Comparison of average number of posts and average time between \sys{} and Baseline in the lab study.}
\label{tab:system_baseline_comparison}
\end{table}

\subsubsection{Post-study questionnaire.}
Post-study questionnaires revealed participants favored \sys{} over the baseline system. Using Wilcoxon signed-rank tests (Bonferroni-corrected $\alpha=0.00714$), \sys{} was perceived to better support engagement with learning materials ($Z=-3.059$, $p=0.0044 < 0.00714$), inspire more ideas during discussions ($Z=-3.059$, $p=0.0031 < 0.00714$), and improve preparation for in-class activities ($Z=-3.059$, $p=0.0066 < 0.00714$) (Figure \ref{fig:study_result1}). Cognitive load metrics showed \sys{} required reasonable amount of mental effort ($p=0.4314 > 0.00714$) and facilitated easy information processing ($p=0.0803 > 0.00714$) (Figure \ref{fig:study_result2}). 

Participants rated all key features of \sys{} positively, particularly its features in valuing inspiring questions, discussion analysis reports, and idea connection (Figure \ref{fig:study_result4}). These results suggest \sys{}'s effectiveness in enhancing learning engagement while reducing cognitive demands.


\subsubsection{Post-study interview.}
\add{We used the reflexive thematic analysis method~\cite{braun2012thematic} for post-study interview analysis}, one of the authors conducted two rounds of open coding to develop an initial codebook. The research team collaboratively reviewed the coded data from the codebook, refining and organizing the results into broader, high-level themes. This process led to the identification of key insights into users' interactions with and experiences of \sys{}.

\textit{KF1: \sys{} enhances more active engagement to discussion.} 
Participants found that \sys{} kept them \textbf{more actively engaged} in online discussions by providing \textbf{quick overviews} of others' posts through \textbf{summaries} and connecting them with relevant peer contributions (P1–6, 8–11). Compared to the baseline system—where users had to \textit{``read everything''} and often \textit{``lost attention''} (P1)—\sys{} offered \textbf{starting points that encouraged participation}. As P6 explained, \textit{``At least I can try and I can start}, whereas without such support, \textit{ I don’t know where I should start''}.

Many participants emphasized that summarization and affinity analysis features \textbf{reduced the effort} needed to find relevant posts. P4 noted they could \textit{``automatically connect me with students that have similar opinions... saving a lot of time''} (P4), while P5 appreciated the \textit{``color coding... to see very quickly at a glance, and then go more in deep''} (P5). P8 similarly valued being able to \textit{``glance through what other students have written without having to read all of it''} (P8).\looseness=-1

Overall, participants agreed that \sys{}'s features lowered the barrier to entry and made it easier to navigate and engage meaningfully in discussions.

\textit{KF2: \sys{} facilitates richer idea inspiration and more meaningful material-based discussion.}
Several participants described the AI in \sys{} as a catalyst or \textit{``third participant''} in the conversation—introducing \textbf{new angles and questions} that extended their own thinking (P4–12). In contrast to the baseline system, where \textit{``when you get stuck, there’s nothing to spark the imagination''} (P11), \sys{} provided prompts that helped students generate ideas. As P12 put it, the inspiring questions and evidence acted as \textit{``an external stimulus... that somehow helps with my reflection,''} (P12) while P4 noted that these prompts were sometimes \textit{``more interesting than my [own] comments''} (P4).

Some participants likened \sys{}’s blending features to having a tutor who highlights \textbf{new directions} and provides \textbf{material-based hints for inquiry}—a especially valuable scaffold in asynchronous settings where immediate feedback is lacking (P8–9, P11–12). Most found the AI-generated inspiring questions and evidences to be high quality, sparking creativity and sustaining meaningful dialogue among peers. Even when not directly applicable, they still \textit{``helped me \textbf{think more}''} and \textit{``\textbf{think critically}''} (P5), fostering more meaningful engagement.


\textit{KF3: \sys{} enhances metacognitive awareness and in-class sessions preparation.}
All participants noted that \sys{}’s learning summary report provided a valuable overview of both the discussion and their own contributions, which many found invaluable for reflection (P1–12). The report helped them to revisit key points and connect them to the material without rereading every post or the full article. As P5 explained, \textit{``I would be able to look at what exactly I wrote and relevant material... before class, in \textbf{a very easy to digest way}, instead of trying to go through everyone’s comments again or have to reread... the really, probably long article”} (P5). P2 added, \textit{``It’s nice for recall to have these keywords that are linked to specific paragraphs”} (P2).

Beyond convenience, participants also appreciated how the report supported metacognitive awareness by consolidating discussion history and surfacing gaps in engagement. P8 shared, \textit{``I really liked the part where it summarized all the things I interacted with... it helped me remember it quickly”} (P8). These features supported metacognitive awareness by helping students \textbf{identify not only what they had learned, but also what they had missed}. P5 remarked, \textit{``It helps to identify which topics I didn’t really look at,”} offering an \textbf{immediate sense of knowledge gaps} (P5), while P1 found it useful for \textit{``self-analysis and knowing what you need to focus on more later.”}

Overall, participants felt better prepared for synchronous in-class sessions with \sys{} after reviewing the report, as it enabled a \textit{``quick''} yet \textit{``comprehensive reflection''} on the online discussion (P6).

\textit{KF4: \sys{} supports cognitive processing without causing information overload.}
Despite introducing many novel features, most of the participants reported that \sys{} supported their cognitive processing without causing additional information overload (P1–3, 7–9, 11–12). As P1 put it, \textit{``All the information was really straightforward''} (P1). The participants consistently agreed that the system's affinity analysis and summarization features \textbf{reduced mental load compared to traditional forums}. As P8 shared, \textit{``Reading the summary made me remember things faster. It helps me focus more on important content instead of jumping everywhere''} (P8). Several students noted that they no longer needed to read every post, which can be infeasible in large classes, as \sys{} surfaced relevant content automatically. \textit{``I didn’t have to read every post. The color and grouping really helped me just pick the ones that matched my interest''} (P9). With \sys{}, the same participant noted that it \textit{``saves me a lot of effort in filtering out comments that are interesting to me''}, even though there was a learning curve. The summarization feature also played a key role in reducing effort during lengthy discussions. As P6 explained, \textit{``With a lot of text, the system helps me \textbf{keep focusing}. I could look at just the highlighted parts or the summary, and I don’t have to scroll through all''} (P6). Overall, students appreciated not having to manually sift through every comment, and found the affinity analysis, inspiring questions and evidences both clear and helpful.\looseness=-1

\textit{User Challenges and Feedback.} Participants identified several challenges they faced when using \sys{}:

First, a recurring concern among participants was the \textbf{inconsistent granularity} of summarizations and affinity analysis. While summaries and affinity analysis aimed to aid interpretation, students found that sometimes, they were either too vague or overly specific to be useful. As P3 noted, \textit{``A lot of the [summarized] comments were just simplifying into keywords that were a bit more broad than I was hoping''} (P3). Features like affinity analysis, though helpful in surfacing diverse perspectives, were also sometimes seen as unreliable or unclear. P9 remarked, \textit{``It's hard to find those similar ideas... It’s not really helped me in finding the relevant posts and comments. It’s not that straightforward''} (P9), and P5 added, \textit{``There are instances where it didn’t accurately capture everything... It’s a good starting point, but you \textbf{need to go deeper} to really understand''} (P5).\looseness=-1

Second, participants also critiqued the rigidity of the conceptual blending framework, which categorized relationships into the \textbf{pre-defined categories}: similarity, contrastive, and complementary. While they are generally useful, this structure sometimes failed to reflect the more complex and context-dependent ways in which students relate ideas. As P6 put it, \textit{``In some cases it makes sense, but some cases it doesn’t make sense''} (P6).

Furthermore, many participants experienced a \textbf{initial learning curve} and cognitive overload when first using the system. Although the interface became more intuitive over time, early use was described as \textbf{mentally taxing}. As P2 explained, \textit{``Some of the functionality is just too overwhelming to me, so I don’t really know how to use it''} (P2).\looseness=-1

\add{Finally, while the system was generally well-received, several potential challenges concerning its impact on \textbf{students’ cognitive engagement and self-regulation} warrant further reflection. First, although most participants noted that the system supported their thinking without replacing critical engagement, some raised concerns that \textbf{less intrinsically motivated} students might become \textbf{overly reliant} on AI-generated summaries, suggested questions, and keywords—thereby reducing opportunities for active, self-directed thinking. As P11 reflected, \textit{``If I'm the teacher, I'm encouraging students to use basic features mostly... If I'm a student, it depends on how much effort I want to put in''} (P11).}

\add{In some cases, the system’s interactive features and visual elements also \textbf{diverted students’ attention from} engaging directly with the \textbf{source material}. For instance, P9 shared, \textit{``I feel like I was just clicking through things... I looked at what the system gave me instead of going back to think about the article'' (P9).}}

\add{Moreover, the visibility of students’ activity and the presence of generated reports introduced a subtle sense of accountability, which, while motivating for some, \textbf{created pressure} for others to \textbf{perform better}. As P12 expressed, \textit{``It makes me feel like I need to be better or say something more useful next time''} (P12).}

\add{These dynamics highlight the importance of balancing system support with learner autonomy and underscore how design choices may shape students’ cognitive engagement and emotional experiences.}

The participants also offered some suggestions for enhancing the system’s functionality. Specifically, they called for more nuanced blending structures that can move beyond the fixed categories to better reflect the complex and layered ways students relate ideas in academic discussions (P3, P4). In parallel, they highlighted the value of personalization and adaptive support, suggesting that the onboarding processes should be tailored to users' prior experience and that the interface should adapt over time to match individual usage patterns (P2, P4). These suggestions reflect a desire for systems that can accommodate diverse cognitive styles and evolving user needs.

\section{Exploratory In-Class Deployment}
We conducted a lightweight deployment of \sys{} in an authentic college-level HCI course in a private U.S. university that utilizes a flipped classroom approach. This deployment was designed as an initial exploration rather than a formal study. The goal was not to conduct a comprehensive evaluation with controlled conditions, but to complement the lab study findings by exploratively observing how students engage with the system in a real educational setting. 

The students in this course read two HCI-relevant articles for each class session, engaged in pre-class asynchronous discussions with peers about the readings, and participated in in-class discussions where they further explored concepts from their pre-class discussions. They typically use the discussion forum in Canvas for the discussion. For the purpose of the deployment, we asked them to use \sys{} for one week for the class session on \textit{Ethics in HCI}, roughly two-thirds into the semester. The two assigned readings were~\cite{Bruckman2014} and~\cite{kapania2024imcategorizingllmproductivity}.

Through this exploratory deployment, our aim was to gain preliminary insight into the following questions.

\begin{itemize}
    \item \textbf{RQ1}:How does \sys{} influence students’ engagement strategies and cognitive processing in real-world flipped classroom settings?
    \item \textbf{RQ2}: How do students perceive the role of AI-assisted features in supporting reflection and preparation in real-world flipped classroom contexts?
    \item \textbf{RQ3}: What role does the reflective report play in supporting classroom participation under high reading demands in real-world flipped classroom settings?
\end{itemize}

\subsection{Participants}
The exploratory deployment included 21 students (7 Female, 14 Male) from a college-level HCI course. The majority (17) were fourth-year senior students, with the remaining participants being master's (2) and doctoral students (2). All participants came from engineering backgrounds, with one student also pursuing a business major.\looseness=-1

Most participants (20) had prior experience with online discussions for academic purposes, with Perusall (17), Canvas (18), and Google Docs (14) being the most commonly used platforms. Regarding pre-class discussion habits, 11 participants reported occasionally discussing learning materials before class, 7 rarely engaged in pre-class discussions, and 3 often discussed materials before most class sessions. Most participants (18) had previous experience with flipped classroom approaches, making them familiar with the pedagogical context of our deployment.

\subsection{Study Procedure}
Students used \sys{} to complete pre-class reading and discussion of two assigned articles. To stimulate meaningful engagement, we pre-populated the system with 30 GPT-4o-generated discussion posts for each article. Students received tutorial materials introducing \sys{}’s core AI-assisted features, including conceptual affinity navigation, content summarization, and discussion reflection tools.\looseness=-1

After familiarizing themselves with the system, students completed their reading and participated in asynchronous discussions through the platform. Following the use of the system, students were invited to complete a questionnaire about their experience. We conducted 60-minute semi-structured interviews with five (detailed information can be found in Table \ref{tableclass} in the Appendix) willing participants to explore their impressions, perceived benefits, and feature-specific feedback. All interviews were conducted via Zoom and recorded with participants' consent.

\subsection{Findings}
\subsubsection{Post-study questionnaire.}
The post-study questionnaire revealed that students self-reported educational benefits from the use of \sys{}. As shown in Figure \ref{fig:study_result5}, participants reported enhanced engagement with course materials, increased ideation during peer discussions, improved reflection on prior learning activities, and greater preparedness for in-class collaboration.   Cognitive load assessments further indicated that students were able to process information efficiently while exerting relatively low mental effort (Figure~\ref{fig:study_result6}), suggesting that \sys{} successfully balanced cognitive support with ease of use.

\subsubsection{Post-study interview.}
Based on the reflexive thematic analysis approach~\cite{braun2012thematic}, we began by having one author conduct one round of open coding to generate a preliminary set of codes. These codes served as the foundation for a structured analysis of the interview data. The entire research team then engaged in a collaborative process to review, refine, and synthesize the coded segments, ultimately distilling them into overarching themes. Through this process, we uncovered core insights into how users engaged with and experienced \sys{}.

\textit{KI1: In-class study confirms and extends prior findings on engagement and cognitive support.}
The in-class study reinforced earlier findings (KF1, KF2, KF4), demonstrating that \sys{} effectively supports engagement, reflection, and cognitive processing. Aligned with KF1, participants valued features such as summarization, affinity-based grouping, and inspiring blends, which enabled \textbf{efficient entry} into discussions and supported \textbf{flexible engagement strategies} (e.g., private reading before discussion, or navigating between learning materials and peer posts based on context).

Consistent with KF2, the system was found to promote more reflective and meaningful contributions. Students noted that \sys{} encouraged deeper thinking and helped them view the discussion as a \textbf{dynamic conversation} rather than isolated replies.

Notably, while the lab study used shorter materials (around 1,000 words each), the in-class setting involved longer readings (around 9,000 words each). Despite the increased complexity, participants reported that \sys{} helped them \textbf{stay focused and oriented}, confirming its effectiveness in supporting situated reading and preparatory thinking by structuring content to reduce cognitive load (KF4).

\textit{KI2: AI features scaffold---not replace---student thinking}
Participants consistently viewed \sys{}'s AI-powered features as \textbf{scaffolds that supported—rather than replaced—their cognitive effort}. Rather than expressing concerns about overreliance, they emphasized how the system helped them process ideas more effectively and sustain meaningful engagement. As P2 noted, \textit{``It’s active learning, but facilitated… I’m making my own mind, but it gives me hints''} (P2). P5 similarly reflected, \textit{``It brings together different threads I didn’t think to connect''} (P5).

AI-generated prompts, such as inspiring questions and linked evidence, were seen as helpful guides that deepened thinking without overwhelming users. P4 explained, \textit{``It kind of helps you start to process everything and make your own opinions... it becomes less just like a one-way reading''} (P4).

Importantly, participants described these features as valuable for classroom preparation. As P5 shared, \textit{``Sometimes I’ve forgotten what I read—this [report and question review] lets me review without rereading everything''} (P5). Across participants, there were \textbf{no concerns about AI hindering learning}; instead, it was seen as a tool that enhanced focus, curiosity, and readiness for class.

\textit{KI3: The learning report supports both cognitive recall and emotional readiness.}
While the lab study showed that \sys{}'s report supported review, the in-class setting revealed its greater value under higher cognitive demands. Unlike the shorter lab texts (around 1,000 words), classroom readings were substantially longer (around 9,000 words), making the report \textbf{essential for managing information overload} and \textbf{preparing for live participation} (P1–5), \add{thereby supporting more active class engagement}.

First, instead of rereading entire texts or scanning scattered comments, students used the report to quickly identify key focus areas and recall prior engagement. \add{As P3 reflected, \textit{``The report brings them all together. I don’t need to go back to each post to recall what I did''} (P3). Students emphasized the easiness and efficiency of using the report in streamlining the review process. P4 said, \textit{``It’s more convenient... I don’t have to click into everything again''} (P4).}

Additionally, students described the report as \textbf{emotionally motivating}, marking the end of the pre-class phase and easing the transition into live discussion. As P2 shared, \textit{``It gives me closure—like a summary of what I’ve done in a game, which is encouraging''} (P2). By supporting both recall and confidence, the report played a key role in fostering active participation—especially in fast-paced classrooms. 

\add{Finally, the report contributed to classroom readiness by helping students arrive more focused and prepared. As P5 noted, \textit{``It shows you what you focused on and what you didn’t... so when there’s a question in class, I kind of know \textbf{what to look for and what I can contribute}''} (P5).} In a conversation with the class instructor, they also reported that the students arrived more prepared, with clearer questions and a better sense of what they want to contribute, making in-class discussions more \textbf{focused, equitable, and student-driven}.
\section{Discussion}
\subsection{Promote Asynchronous, Material-Based Discussion For Flipped Classroom via Conceptual Blending}
Flipped classrooms require active pre-class engagement through content comprehension, reflection, and asynchronous interaction~\cite{kim2014flipped, o2015asynchronous}. Our formative study revealed that students struggle with meaningful participation in these discussions, which limits their in-class preparedness. \looseness=-1

Conceptual Blending Theory~\cite{fauconnier1998blending} offers a framework to address these challenges by explaining how people integrate elements from different mental spaces to produce new understandings~\cite{Fauconnier2002Way}. This process combines disparate knowledge frames~\cite{fauconnier1998conceptual}, enabling analogical reasoning and problem solving~\cite{Veale2019Conceptual}. When applied to material-based discussions, conceptual blending provides cognitive scaffolding that helps students connect their ideas with peers' contributions and course materials, better preparing them for face-to-face learning.

\sys{} exemplifies the use of conceptual blending principles to enhance material-based asynchronous discussions in flipped classrooms. This approach preserves student agency while structuring the blending process---encouraging independent engagement with materials before viewing peers' contributions, visualizing conceptual connections between posts, and generating material-grounded prompts that help students integrate their ideas with peers' perspectives. This approach facilitates the creation of new knowledge structures while maintaining strong connections to course materials.\looseness=-1

\add{While the conceptual affinity navigation and blending features strongly scaffold conceptual integration, it is important to reflect on whether the detailed level of scaffolding provided was entirely necessary. Future work should explore if simpler, lighter-weight approaches—such as basic keyword-based visualizations—might achieve similar learning outcomes without the cognitive demands associated with richer visualizations and interactions.}

Looking forward, AI presents opportunities to extend conceptual blending support in adaptive and disciplinary-specific ways. Future systems could offer dynamic scaffolds that adjust to students’ developing expertise, disciplinary norms, or learning preferences. Additionally, the concept of temporal blending---linking discussions across multiple phases of a course---could help students build longitudinal cognitive bridges, tracking how their ideas evolve over time and reinforcing connections between asynchronous preparation and classroom dialogue.

\subsection{Facilitate the Transition from Self-Learning to In-Class Activities}

Effective flipped classrooms depend on students' pre-class self-learning, yet many struggle to connect this preparation with subsequent in-class activities \cite{bishop2013flipped}. This perceived disconnect between preparatory tasks and classroom learning negatively impacts engagement and motivation \cite{diaz2011blended}. Without structured bridging strategies, students often arrive unprepared or disengaged, undermining the pedagogical benefits of the flipped approach.

Our work in \sys{} demonstrated the potential of using the generation of personalized reports of engagement histories to enhance the metacognitive awareness of students and strengthen the connections between self-learning and classroom activities. These reports help students monitor their engagement, identify knowledge gaps, and recognize relationships between their pre-class work and specific learning outcomes.  By explicitly linking pre-class engagement to in-class outcomes, the system scaffolds both awareness and readiness, which can equip students for more focused, confident participation in class.\looseness=-1

\add{Nevertheless, aspects of these personalized reports could potentially be streamlined. Interactive elements such as detailed peer-interaction charts provide rich analytical insights but simpler visual or textual summaries might also effectively promote metacognition with less complexity. Exploring the optimal balance between interactivity and simplicity represents an important direction for future refinements.}

Findings from our deployment highlight the importance of emotional closure and confidence building in preparing students for classroom dialogue. The system's learning reports helped students not only recall content but also feel cognitively and emotionally ``ready'' for participation---suggesting that tools like \sys{} can serve to reduce anxiety and increase perceived competence in synchronous classroom settings.

Future research could expand beyond personalized reports to create more comprehensive personalizations in flipped learning ecosystems. Future real-time adaptive systems can provide in-moment scaffolding during pre-class learning, adjusting resources and prompts based on detected confusion or engagement patterns. Multi-modal learning supports could incorporate interactive simulations, audio feedback, and visual concept mapping to accommodate diverse learning preferences.

From the instructional side, AI could generate classwide preparation profiles to help instructors anticipate misconceptions, surface emerging themes, and strategically guide in-class discussions or interventions.

Together, these advancements can help move flipped classrooms from content-delivery alternatives to holistic learning environments where AI augments human judgment, fosters deeper reflection, and promotes meaningful collaboration across the boundaries of time and modality.

\subsection{The Role of AI Systems in Supporting and Extending Existing Educational Practices}

As AI systems become increasingly integrated into educational settings, there is growing debate about their appropriate role in the learning process~\cite{zhang2023visar,zhang2022storybuddy}. While many existing tools emphasize automation---such as auto-grading, content recommendation, or conversational tutoring---\sys{} represents a shift toward AI-as-augmentation, where the goal is not to replace instructors or students' cognitive labor but to scaffold and extend existing educational practices such as the flipped classroom model.

Rather than generating answers or directing the learning path, \sys{} functions as a \textit{reflective partner} that aligns with and amplifies the values of learner agency, collaboration, and meaning-making. The system supports essential flipped classroom practices, such as pre-class reading, asynchronous discussion, and in-class participation, not by redesigning pedagogy around AI, but by embedding AI meaningfully within it. This approach reflects a broader vision of AI as an embedded pedagogical agent that enhances, rather than disrupts, established instructional models.

Specifically, the design of \sys{} views AI as a tool for cognitive amplification~\cite{zhang2025ladica}, which helps students engage more deeply with learning materials, peers' ideas, and their own thinking. Features such as conceptual blending, content summarization, and personalized learning reports reduce the cognitive friction of navigating complex discussions and lengthy texts. Importantly, these tools do not dictate meaning or prescribe answers; they prompt students to construct connections and reflectively synthesize diverse inputs. In this model, AI becomes an invisible but powerful support that surfaces connections, highlights gaps, and gently nudges reflection without overtaking the learning process.

\section{Limitation and Future Work}
Despite its promising contributions, the current version of \sys{} has several technical and design limitations that offer opportunities for future enhancement.

First, the system currently relies on fixed prompt templates for affinity analysis, content summarization, and the generation of inspiring questions and evidence. While effective in many cases, these templates may produce outputs with inconsistent levels of granularity depending on the type of reading material or discussion post. This occasionally results in summaries or affinity labels that are either overly general or excessively specific, potentially affecting their interpretability and usefulness.

Second, \sys{} supports three predefined conceptual blending types—similarity, contrastive, and complementary—based on mappings from Conceptual Blending Theory~\cite{fauconnier1998blending}. However, in practice, the ways in which students connect ideas can be more fluid, context-dependent, and structurally complex than these categories capture. Future iterations of \sys{} could explore more adaptive or user-generated blending structures that reflect the nuanced ways students relate and synthesize ideas.

In addition to the implementation limitations of \sys{}, the system leverages GPT-4o\footnote{\href{https://platform.openai.com/docs/models/gpt-4o}{https://platform.openai.com/docs/models/gpt-4o}} and LangChain\footnote{\href{https://www.langchain.com/}{https://www.langchain.com/}} to generate content based on the learning material. While these models are generally robust and produce semantically relevant outputs, generated responses may occasionally lack depth, contain repetition, or misalign with user expectations. These inconsistencies may hinder engagement or introduce frustration. In designing \sys{}, we have positioned AI outputs as cognitive scaffolds rather than authoritative responses—tools meant to stimulate exploration, critical thinking, and deeper engagement rather than provide conclusive answers.\looseness=-1

In terms of study design, several limitations should be acknowledged. 

\add{First, the structured and time-limited nature of our lab study may have influenced participants’ behavior, potentially amplifying engagement due to the “under study” effect. While this setup allowed us to observe focused interactions with the system, it may not fully reflect how students would engage with \sys{} in more naturalistic, unmonitored settings. Future in-the-wild deployments are needed to better assess sustained use and authentic engagement patterns.}

Second, the learning materials used in this study were drawn primarily from university-level courses in the arts, humanities, and social sciences. While these domains align well with the open-ended, discussion-based nature of \sys{}, the platform’s effectiveness in more technical or quantitatively oriented disciplines---such as mathematics, engineering, or physics---remains to be investigated. Future deployment studies are needed to evaluate the applicability of \sys{} across a wider range of subject areas and instructional formats.\looseness=-1

Additionally, the participant cohort in our exploratory deployment was relatively homogeneous, consisting primarily of senior undergraduate and graduate students from engineering backgrounds enrolled in a single class at a high-ranking US private institution. Expanding participant demographics in future deployment studies will help assess how \sys{} supports students with varying levels of academic experience, disciplinary perspectives, and learning preferences. \looseness=-1

\add{Finally, the sample sizes of participants in our formative (n=4) and lab studies (n=12) were relatively small. However, the formative study included students from different academic stages and disciplines, offering a preliminary understanding of learner needs. The lab study involved participants from engineering, social sciences, and natural sciences, allowing us to observe how \sys{} supports diverse reasoning styles and discussion practices. Future work could explore large-scale deployments to assess \sys{}’s effectiveness and adaptability in varied learning contexts.}

Looking ahead, we also plan to extend \sys{} to support multimodal learning materials, including video and audio content. This enhancement would enable students to integrate discussions with diverse formats, broadening \sys{}’s applicability to flipped classrooms, hybrid learning environments, and courses that rely on non-textual instructional resources. By enabling flexible engagement with multimodal content and fostering richer peer-to-peer synthesis, our goal is to support deeper understanding, promote multi-perspective thinking, and cultivate more inclusive, collaborative learning environments.

\section{Conclusion}
This paper introduced \sys{}, an AI-assisted discussion platform designed to enhance pre-class learning experiences in flipped classrooms. Our research identified several challenges students face during asynchronous online discussions, including difficulties engaging with temporally dispersed contributions, connecting discussions with learning materials, and developing metacognitive awareness. In response, \sys{} offers integrated cognitive and metacognitive support to (1) help students identify conceptually related posts, (2) scaffold knowledge integration through conceptual blending, (3) promote deeper engagement with peer contributions, and (4) enhance metacognitive awareness through structured reflection. 

Our within-subjects lab study (n=12) demonstrated that participants effectively used \sys{} for pre-class learning and found the system valuable in improving discussion engagement, generating new ideas, supporting reflection processes, and increasing preparedness for in-class collaborative activities. These findings suggest that \sys{} can address key challenges in flipped classroom environments while supporting important cognitive and metacognitive processes essential for effective learning.

\begin{acks}

This work was supported in part by a Notre Dame–IBM Technology Ethics Lab Award, an NVIDIA Academic Hardware Grant, a Google Research Scholar Award, a gift from Adobe, NSF grant DRL-2437113, and a University of Notre Dame Strategic Framework Grant. We thank Xinyue Chen, Yuwen Lu, and Ruotong Wang for their valuable feedback on the \sys{} project, and Shuyu Wang for insightful suggestions on the teaser figure and demo video. We are also grateful to all study participants and collaborators for their time and contributions.

\end{acks}

\bibliographystyle{ACM-Reference-Format}
\bibliography{ref}

\clearpage
\begin{table*}[]
\section*{APPENDIX}
\vspace{0.2cm}
\centering
\renewcommand{\arraystretch}{1.2} 
\begin{tabular}{ccccccccc}
\hline
\textbf{PID} & \textbf{Gender} & \textbf{Educational Background} &  
\textbf{Academic Background}
\\ 
\hline
1  & Female & Undeclared & Undergraduate    \\ 
2  & Female & Engineering& Undergraduate \\
3  & Female & Biochemistry& Undergraduate \\
4  & Female & Engineering& Graduate \\
\hline
\end{tabular}
\vspace{0.2cm}
\caption{We recruited four participants with diverse educational backgrounds and prior experience using online learning platforms that support asynchronous discussions in flipped learning environments.}
\label{tableformative}
\end{table*}
\begin{table*}[htbp] 
\centering
\renewcommand{\arraystretch}{1.2} 
\begin{tabular}{ccccccccc}
\hline
\textbf{PID} & \textbf{Gender} & \textbf{Academic Background} & \textbf{Discussion Tool Usage Exp.}    \\ 
\hline
1  & Male & Engineering & Perusall   \\ 
2  & Female & Engineering & Canvas, Google Docs, Perusall \\
3  & Female & Engineering & Canvas, Google Docs, Perusall \\
4  & Female & Engineering & Canvas, Notion, Google Docs, Perusall \\
5  & Female & Engineering & Canvas, Perusall         \\
\hline
\end{tabular}
\vspace{0.2cm}
\caption{We conducted interviews with 5 students from the class who had used \sys{} over the span of one week.}
\label{tableclass}
\end{table*}

\begin{table*}[htbp] 
\centering
\renewcommand{\arraystretch}{1.2} 
\begin{tabular}{ccccccccc}
\hline
\textbf{PID} & \textbf{Gender} & \textbf{Academic Background} & \textbf{Discussion Tool Usage Exp.}    \\ 
\hline
1  & Female & Engineering & Piazza, Canvas, Google Docs, Perusall   \\ 
2  & Female & Engineering & Piazza, Canvas, Notion, Google Docs \\
3  & Male & Engineering & Piazza, Canvas, Notion, Google Docs \\
4  & Male & Engineering & Canvas, Notion, Google Docs \\
5  & Female & Social Sciences & Piazza, Canvas, Google Docs, Perusall \\
6  & Female & Engineering & Google Docs \\
7  & Female & Social Sciences & Notion, Google Docs \\
8  & Female & Engineering & Piazza, Canvas, Google Docs \\
9  & Male & Natural Sciences & Piazza, Canvas, Notion, Google Docs \\
10 & Male & Engineering & Piazza, Notion, Google Docs\\
11 & Male & Natural Sciences & Piazza, Canvas, Google Docs, Perusall \\
12 & Female & Engineering & Piazza, Canvas, Notion, Google Docs \\
\hline
\end{tabular}
\vspace{0.2cm}
\caption{We recruited 12 participants with different academic background and have prior experience with asynchronous discussion tools.}
\label{tablelab}
\end{table*}
\begin{table*}[h]
\centering
\begin{tabular}{|p{0.2\textwidth}|p{0.75\textwidth}|}
\hline
\textbf{Task} & \textbf{Prompt Template} \\
\hline
Affinity Navigation & 
Analyze the relationship between a primary discussion card and a collection of related cards.

1. Relevance Analysis:
   - Compare the content of the primary card with each card in the collection
   - Calculate relevance scores
   - Generate a ranked order based on content relevance
   - Ensure the primary card's original position is preserved at the top

2. Shared affinity types Identification:
   - For each comparison, identify a affinity type (1-2 words) that captures the conceptual relationship
   - Affinity type should reflect the nature of the connection between cards
   - Use "none" if no meaningful relationship is found

3. Relevance Classification:
   - Categorize relationships as: high, medium, or low
   - Assign percentage scores to indicate relative strength
   - Provide specific themes for each relationship \\
\hline
Inspiring Question & 
Generate a thought-provoking discussion question based on the provided content, keywords, and their descriptions. The question should align with one of these discussion styles:

Discussion Styles:
1. Similarity Focus
   - Encourage students to identify shared perspectives
   - Help them build on common ground
   - Foster collaborative thinking

2. Contrastive Focus
   - Promote respectful debate of different viewpoints
   - Encourage critical analysis of opposing arguments
   - Develop skills in defending positions

3. Complementary Focus
   - Guide students to find ways ideas can enhance each other
   - Identify how different perspectives fill knowledge gaps
   - Explore how combining viewpoints creates deeper understanding

Requirements:
- Question should be clear and engaging
- Length: 20-30 words
- Must directly relate to provided keywords and content
- Should naturally lead to the specified style of discussion \\
\hline
Material-grounded evidence & 
Identify three pieces of evidence from the article to help students engage with the discussion question.

For each piece of evidence:
1. Extract a key concept (1-2 words) that connects the evidence to the question
2. Provide the exact text from the article that supports this concept
3. Maintain all original formatting, including punctuation and capitalization

Purpose:
- Support students in developing well-reasoned responses to the discussion question
- Help students connect specific parts of the text to their arguments
- Enable evidence-based participation in class discussions \\
\hline
Discussion overview & 
Analyze the provided materials which include:
1. Original articles
2. Student discussions and posts about these articles
3. Individual user's comments

Your analysis should:

1. Discussion Topic Analysis:
   - Identify key discussion topics and convert them to 1-2 word keywords
   - Summarize each user's specific contributions under these topics
   - Provide strategic suggestions for deepening these discussions

2. Individual Engagement Analysis:
   - Review the user's comments to identify:
     a) Topics of High Engagement: Extract keywords from sections with active participation
     b) Topics of Low Engagement: Identify keywords for less-engaged sections

3. Community Focus Analysis:
   - Identify "hotSpot" sections that generated significant discussion
   - For each hotSpot: Create concise keywords (1-2 words) \\
\hline
Discussion Analysis & 
Analyze a student's discussion contribution for a specific paragraph in the article.

Input provided:
1. The student's comment on a specific paragraph and paragraph index
2. The original paragraph text and its index

Please provide:
1. Keywords Analysis (1-3 keywords)
2. Discussion Summary: Start with "You discussed..." (limit: 30 words)
3. Engagement Suggestions: Start with "You could..." (limit: 30 words) \\
\hline
Multi-framework keyword highlighting & 
Analyze two discussion cards to recommend the most effective discussion approach.

Task:
1. Discussion Style Analysis:
   - Calculate percentage distribution across three potential discussion styles:
     * Similarity-based, Contrastive, and Complementary Discussion
   - Ensure percentages total 100

2. Evidence Selection:
   - For each relationship type with non-zero percentage:
     * Extract brief quotes (1-3 words) from original text
     * Select from card1.content and card2.content only

3. Discussion Direction:
   - For each non-zero relationship:
     * Provide a discussion aspect (1-10 words) \\
\hline
Content summarization & 
Generate a concise summary of the nested object's content.

Task:
Create 1-3 bullet-point summaries that:
- Capture the main ideas from the object's content field
- Each summary should not exceed 30 words
- Provide fewer points for shorter content
- Generate fresh summaries when regeneration is requested \\
\hline
Key aspect extraction for blending & 
Summarize three keywords and their extended descriptions of two nested objects based on the article.
- Keywords should be 1-2 words
- Extended descriptions should not exceed 20 words
- Each keyword should have corresponding original text in card1 and card2
- Original text must be in card1.content and card2.content, not in children
- Preserve original formatting including punctuation and capitalization \\
\hline
\end{tabular}
\parbox{\textwidth}{\centering
\vspace{0.2cm}
\caption{LLM Prompt Templates Used in the Key Features of \sys{}.}
}
\label{tab:prompt_templates}
\end{table*}

\end{document}